\documentclass[traditabstract]{aa}

\usepackage{graphicx}
\usepackage{txfonts}
\usepackage{natbib}
\usepackage[colorlinks=true,citecolor=blue]{hyperref}
\usepackage{xspace}
\usepackage{array}
\usepackage[labelformat=simple]{caption,subcaption}
\usepackage{amsmath,amssymb}
\usepackage{bm}
\usepackage{gensymb}
\usepackage{color}

\bibpunct{(}{)}{;}{a}{}{,}

\newcommand{\mic}{$\mu$m\xspace}
\newcommand{\as}{\hbox{$^{\prime\prime}$}\xspace}
\newcommand{\lsd}{\hbox{$\lambda/D$}\xspace}

\newcommand{\degre}{\degree\xspace}

\newcommand{\C}{\ensuremath{\mathcal{C}}\xspace}
\renewcommand{\S}{\ensuremath{\mathcal{S}}\xspace}
\newcommand{\F}[1]{\ensuremath{\mathcal{F}}{#1}\xspace}
\newcommand{\I}{\ensuremath{\mathcal{I}}\xspace}

\begin{document}

\title{Validation of strategies for coupling exoplanet PSFs into single-mode fibres for high-dispersion coronagraphy}

\titlerunning{Validation of strategies for coupling PSFs into single-mode fibres}

\author{
  M. El Morsy\inst{\ref{lam}} 
  \and
  A. Vigan\inst{\ref{lam}} 
  \and
  M. Lopez\inst{\ref{lam}} 
  \and
  G.P.P.L. Otten\inst{\ref{lam}, \ref{as}} 
  \and
  E. Choquet\inst{\ref{lam}} 
  \and
  F. Madec\inst{\ref{lam}} 
  \and
  A. Costille\inst{\ref{lam}} 
  \and
  J.-F. Sauvage\inst{ \ref{onera}, \ref{lam}}
  \and \\
  K. Dohlen\inst{\ref{lam}} 
  \and
  E. Muslimov\inst{\ref{Nova}, \ref{lam}, \ref{kazan}} 
  \and
  R. Pourcelot\inst{\ref{lam}, \ref{oca}}
  \and
  J. Floriot\inst{\ref{lam}} 
  \and
  J.-A. Benedetti\inst{\ref{lam}} 
  \and
  P. Blanchard\inst{\ref{lam}} 
  \and
  P. Balard\inst{\ref{lam}} 
  \and
  G. Murray\inst{\ref{dh}} 
}

\institute{
    Aix Marseille Univ, CNRS, CNES, LAM, Marseille, France \label{lam}
    \\ \email{\href{mailto:mona.elmorsy@lam.fr}{mona.elmorsy@lam.fr}}
    \and
    Universit\'e C\^ote d'Azur, Observatoire de la C\^ote d'Azur, CNRS, Laboratoire Lagrange, France \label{oca}
    \and
    Academia Sinica, Institute of Astronomy and Astrophysics, 11F Astronomy-Mathematics Building, NTU/AS campus, No. 1, Section 4, Roosevelt Rd., Taipei 10617, Taiwan \label{as}
    \and
    Center for Advanced Instrumentation, Durham University, Durham, DH1 3LE, UK
    \label{dh}
    \and
    DOTA, ONERA, Université Paris Saclay (COmUE)
    \label{onera}
    \and
    NOVA Optical IR Instrumentation Group at ASTRON Oude Hoogeveensedijk 4, 7991 PD Dwingeloo, The Netherlands
    \label{Nova}
    \and
    Kazan National Research Technical University named after A.N. Tupolev KAI, 10 K. Marx, Kazan, Russia, 420111
    \label{kazan}
}
\date{Received 24 February 2022; accepted 09 May 2022}

\abstract
{On large ground-based telescopes, the combination of extreme adaptive optics (ExAO) and coronagraphy with high-dispersion spectroscopy (HDS), sometimes referred to as high-dispersion coronagraphy (HDC), is starting to emerge as a powerful technique for the direct characterisation of giant exoplanets. The high spectral resolution not only brings a major gain in terms of accessible spectral features, but also enables a better separation of the stellar and planetary signals. Ongoing projects such as Keck/KPIC, Subaru/REACH, and VLT/HiRISE  base their observing strategy on the use of a few science fibres, one of which is dedicated to sampling the planet’s signal, while the others sample the residual starlight in the speckle field. The main challenge in this approach is to blindly centre the planet's point spread function (PSF) accurately on the science fibre, with an accuracy of less than 0.1\,\lsd to maximise the coupling efficiency. In the context of the HiRISE project, three possible centring strategies are foreseen, either based on retro-injecting calibration fibres to localise the position of the science fibre or based on a dedicated centring fibre. We implemented these three approaches, and we compared their centring accuracy using an upgraded setup of the MITHiC high-contrast imaging testbed, which is similar to the setup that will be adopted in HiRISE. Our results demonstrate that reaching a specification accuracy of 0.1\,\lsd is extremely challenging regardless of the chosen centring strategy. It requires a high level of accuracy at every step of the centring procedure, which can be reached with very stable instruments. We  studied the contributors to the centring error in the case of MITHiC and we propose a quantification for some of the most impacting terms.
}


\maketitle
\section{Introduction}
\label{sec:introduction}
Over the past decades, numerous exoplanets have been detected beyond the Solar System by indirect or direct methods allowing us to target different populations and retrieving physical key properties \citep{mayor1995jupiter}. One of the main advantages of directly imaging exoplanets is that it allows quantifying the chemical compositions and abundances of their atmospheres \citep{snellen2014fast, brogi2019retrieving}, and establishing their formation and evolution mechanisms \citep{crepp2012dynamical, piso2016role}. High-contrast imaging (HCI) focuses  on the direct detection of young (<300 Myr) giant (>1 Mjup) exoplanets orbiting at relatively large orbital separation (>10\,ua) from their host star. The high contrast (>10 mag) and small angular separation (<1\as) of these planets with respect to their host star are a severe limitation to their detection, but especially to their atmospheric characterisation. 

High-contrast imaging combines extreme adaptive optics \citep[ExAO, e.g., ][]{fusco2006high} to measure and compensate for the atmospheric distortion in ground-based observations, with coronagraphy to attenuate the on-axis starlight \citep{guyon2006theoretical, Mawet2012}. However, the technique is mainly limited by quasi-static speckles in the focal plane \citep{soummer2007speckle}, which cover the planetary signal. To deal with the challenges of HCI, a new generation of instruments has been implemented on large ground-based telescopes in the last decade: VLT/SPHERE \citep{beuzit2019sphere}, Gemini/GPI \citep{macintosh2014first}, and Subaru/SCExAO \citep{jovanovic2015subaru}. However, the low spectral resolutions of these instruments, typically $R = 50$ for the SPHERE IFS \citep{zurlo2014}, are a major limitation for exoplanet characterisation.

Simultaneously to HCI, high-dispersion spectroscopy (HDS) has proven to be a powerful technique to characterise the atmosphere of exoplanets  using infrared spectrographs such as Keck/NIRSPEC \citep{mclean1998design} or VLT/CRIRES \citep{kaeufl2004}. HDS in the near-infrared (NIR) for ground-based observation has provided a great deal of  essential information, such as the wind speed or the rotational velocity of some transiting exoplanets \citep{snellen2010orbital}. It has also been  used to characterise exoplanets by spectrally resolving molecular features of species present in the atmosphere \citep{brogi2013detection, kok2013detection}. The combination of high angular resolution using AO to first spatially separate the planet from the stellar signal and the implementation of medium- to high-resolution spectroscopy to spectrally filter the stellar light from the planet's light was first   successfully demonstrated by using the VLT/CRIRES instrument with a resolution of $R = 100\,000$ in the K band. Using the MACAO AO system \citep{2003arsenault}, CRIRES was   used to detect the CO lines at 2.3~\mic in $\beta$ Pictoris b and to measure its orbital and rotational velocities \citep{snellen2014fast}. However, these results were partly limited by the relatively poor high-contrast performance of the CRIRES system, which was not   designed for this purpose.

The proper combination of HCI and HDS was first proposed by \citet{riaud2007improving} to improve the sensitivity in contrast to exoplanets, and a more in-depth study was performed by \citet{2015snellen} in the framework of ELT instrumentation. In theory, for ground-based observations  HCI and HDS can respectively reach a planet--star contrast down to $10^{-3}$ and $10^{-4}$, so combining the two to perform high-dispersion coronagraphy (HDC) could theoretically enable  contrasts of $10^{-7}$ or better to be reached  \citep{2015snellen}. More recently, the study by \citet{2017wang} developed a detailed model of HDC observations and analysed the trade-off between starlight suppression and spectral resolution for the characterisation of young giant exoplanets in the NIR. This framework  opened the path for HDC, which is now implemented in several projects on existing telescopes: Keck/KPIC \citep{2021delorme}, Subaru/REACH \citep{kotani2020reach}, and VLT/HiRISE \citep{vigan2018bringing}.

The three projects mentioned above consist in coupling existing high-contrast instruments with medium- or high-resolution spectrographs. They propose  sampling the planetary signal with a single-mode fibre (SMF) and feeding it to a diffraction-limited spectrograph \citep{1994coude, jovanovic2017efficient}. For precision and stability purposes, SMFs are optimal for feeding light into spectrographs \citep{crepp2014improving, 2016jovanovic}. They are used in the focal plane of telescopes for their spatial filtering effect \citep{ghasempour2012single}. However, efficiently coupling a point spread function (PSF) into a SMF is inherently challenging due to their intrinsic properties. A SMF is designated only to support a quasi-Gaussian fundamental mode (LP01). According to the field theory of guided waves, optimal coupling efficiency occurs when the overlap integral between the electric field of the input beam and the LP01 mode is  maximised \citep{1983jeunhomme, 1988neumann, barrell1979optical}. It implies that the complex amplitude of the transverse component of the electric field at the telescope focal plane must match the LP01 mode of the SMF \citep{1993coude}. In the ideal case of an unobstructed pupil, an Airy pattern is formed in the focal plane and is composed of a constant phase core surrounded by successive dark and bright concentric rings where the phase flips by $\pi$. Coupling an Airy pattern into the LP01 mode is bounded by the opposite phase of the first ring, which leads to destructive interference and reduces the coupling efficiency. The maximum coupling efficiency of an Airy pattern into a SMF occurs when only the core matches the quasi-Gaussian mode. An unobstructed circular aperture’s theoretical maximum coupling efficiency into a SMF is $\sim$81\% \citep{Shaklan:88}.

However, most ground-based telescopes have obstructed circular apertures and spiders. The presence of a central obstruction and spiders in the pupil have a significant impact on the coupling efficiency. 
Due to their respective pupils, the Gemini telescopes offers a maximum coupling efficiency of $\sim$70\%, the Keck and the Subaru telescopes $\sim$60\% \citep{jovanovic2017efficient}, and the VLT $\sim$73\% \citep{otten2021direct}. Ground-based observations are also limited by optical aberrations, which are classified into two categories: quasi-static aberrations caused by optical defects and misalignments  \citep[e.g.][]{n2013calibration} and atmospheric aberrations caused by the turbulent atmosphere. Both types of aberrations disturb the wavefront and require the use of an AO system to  re-establish a wavefront that is as flat as possible, which will better couple into a SMF. 
An ExAO correction has proven to increase the Strehl ratio, minimise the aberrations, and therefore bring the injection efficiency closer to its theoretical maximum \citep{jovanovic2017efficient}. 

Finally, in order to maximise the coupling efficiency into a SMF, the telescope PSF must be centred on the fibre. Any offset affects the coupling. \citet{otten2021direct} demonstrate the need for HiRISE to achieve a coupling efficiency >95\% (according to section 3.3 paragraph 2 of the present paper) of the maximum possible to spectrally characterise exoplanets. To reach this goal, the PSF must be centred with the core of the fibre within 0.1\,\lsd. We can tolerate a coupling efficiency as low as 59\% of the maximum, which corresponds to a misalignment between the PSF and the fibre of 0.2\,\lsd. In this scenario we will not be able to characterise our most challenging targets and will need more telescope time to spectrally characterise the others.

At the VLT, two flagship instruments are installed on the Nasmyth platforms of the unit telescope 3 (UT3): the exoplanet imager SPHERE on one side, and the NIR high-resolution spectrograph CRIRES+ \citep{dorn2016+} on the opposite side. The  High-resolution imaging and spectroscopy of exoplanets (HiRISE) project involves implementing a fibre coupling between these two instruments to enable the characterisation of known directly imaged young giant exoplanets at spectral resolutions up to $R = 100\,000$ in the H band. HiRISE is composed of three independent parts: (i) the fibre injection module (FIM) that will be implemented in the IFS arm of SPHERE, (ii) the fibre bundle (FB) that will link SPHERE and CRIRES+, and (iii) the fibre extraction module (FEM) that will be installed on the calibration stage of CRIRES+. HiRISE will benefit from the SPHERE infrastructure for the HCI and from CRIRES+ in NIR for the HDS techniques.

One of the challenges of coupling SPHERE and CRIRES+ resides in successfully injecting the signal of a previously known exoplanet detected by SPHERE into a SMF. The FIM will play a critical role during observations since it will pick up the planet's PSF in SPHERE and inject it into a SMF located in the FB. To maximise the coupling efficiency, the planet's PSF must ideally be positioned with an accuracy better than 0.1\,\lsd (which corresponds to 4~mas in the H band for SPHERE) to minimise the coupling efficiency losses \citep{otten2021direct}. Therefore, it is essential to define the best possible strategy to centre the planet's PSF on a SMF. 

In this paper three different strategies are investigated to fulfil the required centring specification for HiRISE. To explore these strategies we used an upgraded version of the Marseille Imaging Testbed for High-Contrast (MITHiC) \citep{pourcelot2021calibration}. In Sect.~\ref{sec:hirise} we present a brief overview of HiRISE implementation on the VLT and we describe the PSF centring strategy principles. In Sect.~\ref{sec:mithic} we present the MITHiC testbed and detail the description of the FIM. We also explain the limitations of the setup and the specifications required for the centring strategies. In Sect.~\ref{sec:lab_centring_strat} we introduce the description of the centring strategy procedures in laboratory, and we discuss the performance results. Finally, in Sect.~\ref{sec:ccl} we conclude and propose some perspectives.

\section{Centring strategies for HDC}
\label{sec:hirise}
\subsection{Description of HiRISE}

\begin{figure}
    \centering
    \includegraphics[width=0.5\textwidth]{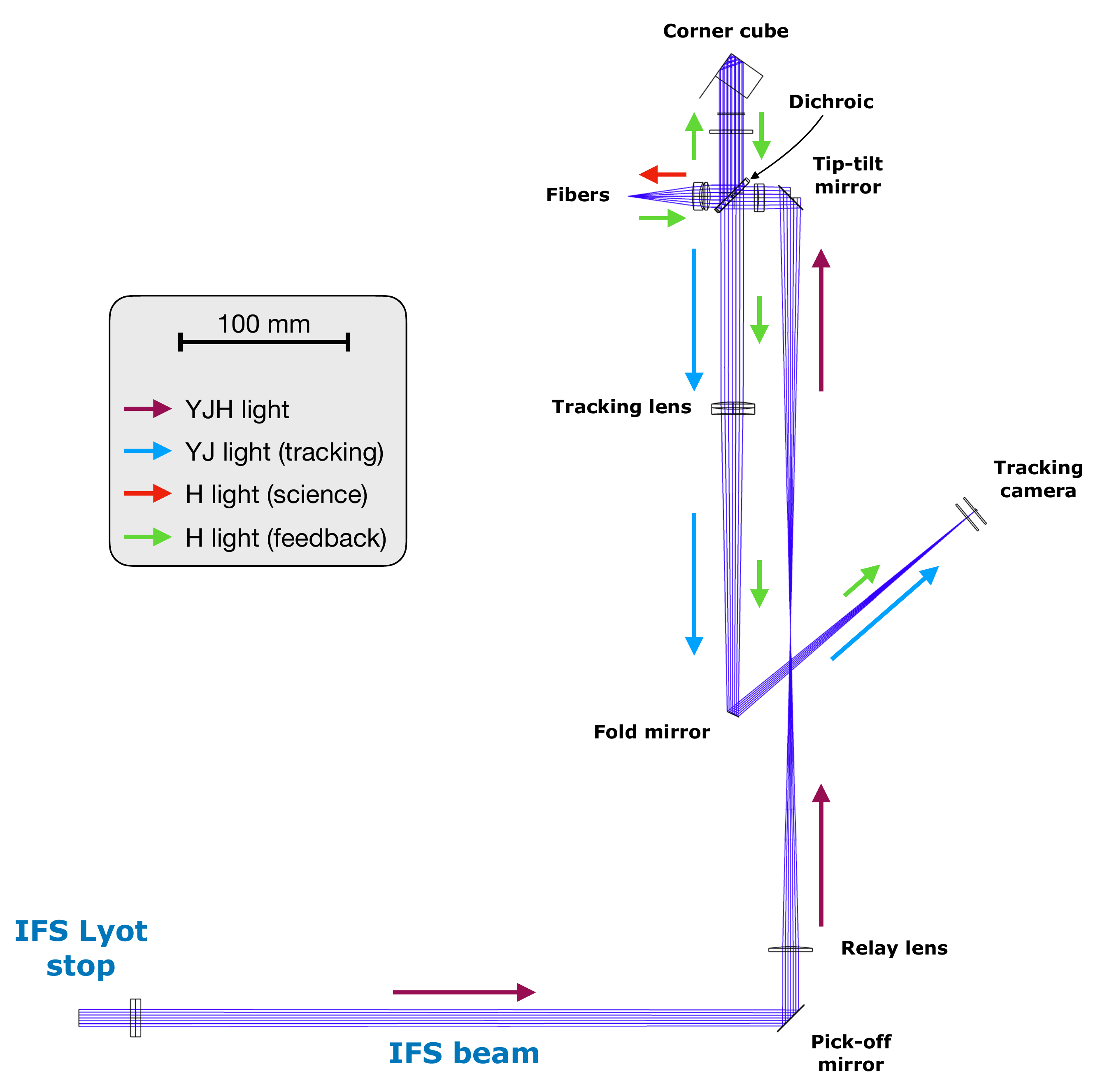}
    \caption{Schematic of the design of the HiRISE/FIM. The main optical components are labelled, as are the photon sharing between wavelengths for the tracking (0.95-1.3\,\mic) and the science (1.4-1.8\,\mic).}
\end{figure}

Conceptually, HiRISE diverts the SPHERE/IFS beam downstream of the ExAO and coronagraph, and injects the light of a known exoplanet into a SMF. This fibre is connected to the NIR spectrograph CRIRES+ on the opposite Nasmyth platform, which is used to disperse the light of the planet at high spectral resolution. The practical implementation requires three inter-connected elements: the FIM implemented inside of SPHERE that takes care of properly injecting the planetary signal into the science fibre, the FB around the telescope that routes the science fibre and additional fibres between the two Nasmyth platforms, and the FEM implemented inside of CRIRES+ that reimages the science fibre's output at the entrance of the spectrograph with the right optical properties. Since the current work focuses on the strategies to best inject the planet's signal into the science fibre, we  describe in greater detail the FIM, and refer the reader to other works for the descriptions of the other elements \citep{vigan2018bringing,otten2021direct}.

The FIM is implemented as a vertical bench immediately downstream the Lyot stop wheel in the IFS arm of SPHERE \citep{beuzit2019sphere}. It includes a pick-off mirror that can be inserted into the IFS beam to send the light towards the FIM instead of the IFS. The SPHERE pupil is reimaged onto a mirror mounted on a S335 piezo tip-tilt platform from \emph{Physik Instrumente}. The beam is then recollimated using a custom achromatic doublet and goes through a dichroic filter that reflects the short wavelengths (0.95-1.3\,\mic; YJ bands) into the `tracking branch' and transmits the longer ones (1.4--1.8\,\mic; H band) to the `injection branch'.

In the tracking branch a focal plane image is formed at F/40 on a C-RED 2 camera from \emph{First Light Imaging} \citep{2020CRED}, and we  refer to it as the tracking camera in this section. This image is sampled at 12.25 mas/pix, which corresponds to Nyquist-sampling at 0.95\,\mic. Although the images from the tracking camera are not expected to be used for any astrophysical interpretation, they will nonetheless be of sufficiently high quality to enable basic processing. In the injection branch, the beam is refocused with a very fast doublet at F/3.5, which has been designed to match the numerical aperture ($NA=0.16$) of the science fibres used in the bundle. The optical system is designed to be telecentric in the injection branch to maximise the injection efficiency even when considering fibres located off-axis. Although the science fibre will be located on-axis to maximise the wavefront quality and therefore the injection efficiency \citep{jovanovic2017efficient}, three additional fibres will be located off-axis to sample the stellar light in the speckle field. In practice, the FB remains static after the alignment of the system, and the focal plane image is moved with respect to the science fibre using the tip-tilt mirror, so as to place the PSF of a known companion onto the fibre. The strategies to centre the PSF accurately on the science fibre are detailed in the following section.

In addition to the fibres dedicated to science, the bundle will also include `feedback fibres', which are connected to a calibration source at $\sim$1.3\,\mic and used to retro-inject signal into the system. Because of the orientation of the dichroic filter, the retro-injected signal is reflected on the dichroic, then on a dedicated corner cube, and then transmitted through the dichroic into the tracking branch. The outputs of the feedback fibres are finally reimaged on the tracking camera superimposed on the focal plane coronagraphic image. The position of the feedback fibres with respect to the science fibre are  calibrated, giving the possibility to use them for accurate centring on the science fibre. This solution was previously proposed to optimise the centring on the science fibre in a situation where the planet's signal remains undetectable in short exposures \citep{mawet2017}. The wavelength of the feedback fibres was selected to be in the transition zone of the dichroic filter because the retro-injected beam has to go through that filter twice. At 1.3\,\mic, the transmission of the dichroic is $\sim$50\%, which means that only 25\% of the photons will ultimately reach the tracking camera.

\subsection{PSF centring strategies on the science fibre}

Immediately after the FIM receives the ExAO-corrected beam from SPHERE it becomes entirely independent from SPHERE and the planet acquisition relies only on the FIM. It consists in centring the planet's PSF on the science fibre with the required accuracy mentioned previously. Since the planet's PSF will not usually be visible on the tracking camera image because of the S/N, the planet acquisition will be performed blindly based on its known relative astrometry with respect to the star and on calibrations performed during the day. Three strategies are foreseen for centring the planet's PSF on the science fibre, and for each strategy, two configurations are possible depending on the presence (or lack) of a coronagraph. For the practical implementation of the centring strategies, a certain number of key calibrations are necessary and are   detailed below.

Several aspects have driven the choice to only investigate these three centring strategies and select the best one to implement for HiRISE. First, they are relatively simple to implement and require only a limited amount of dedicated hardware (either fibres or dedicated sensors). They are then usable from the operational point of view at the telescope. One of the important aspects of HiRISE is that once the system is installed, it should be operable remotely like any VLT instrument. It will not be possible to change or plug any calibration hardware either during the day or the night. And finally, at least one of them has already been demonstrated to work on-sky, which  is an important aspect.

\begin{figure*}
    \centering
    \includegraphics[width=1\textwidth, angle=0]{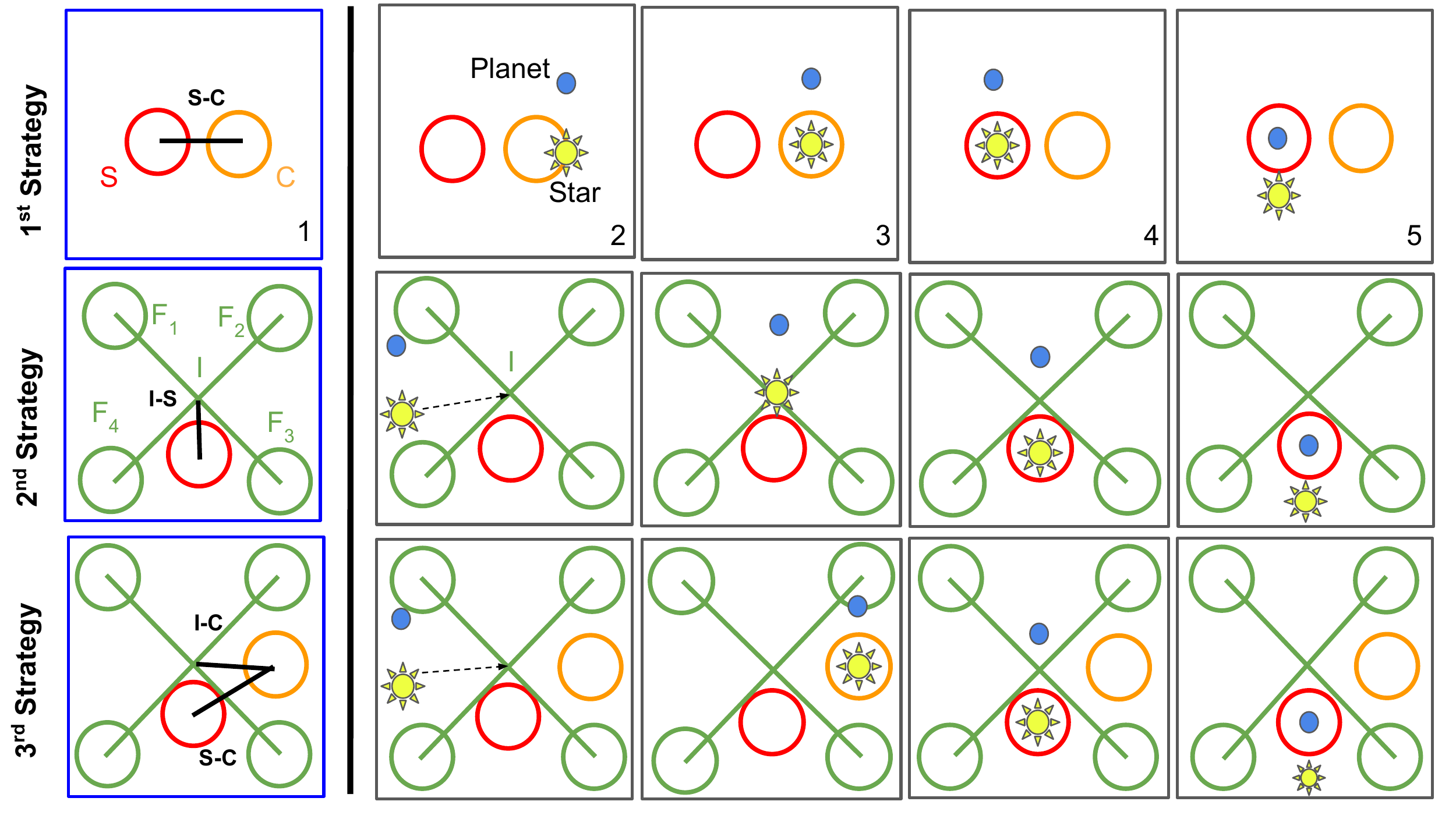}
    \caption{Centring strategy schemes. Each row represents a strategy. The blue frames represent the calibration procedure for each strategy, while the black frames are assigned to the centring procedure. The panels labelled 1--5 show the steps followed to place the planet's PSF into the science fibre.
    \label{fig:strategy_scheme}}
\end{figure*}

The first strategy (Fig.~\ref{fig:strategy_scheme}, top row) relies only on a dedicated centring fibre located close to the science fibre. In the non-coronagraphic configuration the first step consists in placing the stellar PSF on top of the centring fibre to refine the position with an optimisation algorithm that maximises the coupling efficiency of the star into the centring fibre (Fig.~\ref{fig:strategy_scheme}, panel 2). Then, by using the offset pre-calibrated during the laboratory validation of the instrument (Fig.~\ref{fig:strategy_scheme}, panel 1), we place the star onto the science fibre (Fig.~\ref{fig:strategy_scheme}, panel 4). And finally, by using the interaction matrix and using the known right ascension and declination (RA and  DEC) offsets, we apply an offset to the tip-tilt mirror to move the planet on the science fibre (Fig.~\ref{fig:strategy_scheme}, panel 5).

The second strategy (Fig.~\ref{fig:strategy_scheme}, middle row), also in the same non-coronagraphic configuration, uses the science fibre and four feedback fibres that retro-inject light into the system via the dichroic and the corner cube. The four fibres are re-imaged on the tracking camera on top of the stellar PSF. Their diagonal intersection (\I)  in theory provides the position of the science fibre with respect to the focal plane image. Consequently, the first step of the procedure consists in switching on the four feedback fibres to acquire an image with the tracking camera, find the centre of the feedback spots and then determine the position of the science fibre (Fig.~\ref{fig:strategy_scheme}, panel 2). However, an offset may occur between \I and the position of the science fibre. It is due to an alignment mismatch between the science branch and the retro-injection branch, and would need to be calibrated beforehand during the laboratory validation of the instrument (Fig.~\ref{fig:strategy_scheme}, panel 1). Then we position the stellar PSF onto the science fibre by moving the tip-tilt mirror (Fig.~\ref{fig:strategy_scheme}, panel 4) and, as described for the first strategy, we apply an offset to move the planet into the science fibre (Fig.~\ref{fig:strategy_scheme}, panel 5).

The third strategy (Fig.~\ref{fig:strategy_scheme}, bottom row) combines the first and the second strategies using the four feedback fibres, the centring, and the science fibre. The first step consists in switching on the feedback fibres to find the centre of the four spots and the intersection \I (Fig.~\ref{fig:strategy_scheme}, panel 2). We use the tip-tilt mirror to place the star on the centring fibre (Fig.~\ref{fig:strategy_scheme}, panel 3). By using the offset pre-calibrated during the laboratory validation of the instrument (Fig.~\ref{fig:strategy_scheme}, panel 1), we place the star onto the science fibre (Fig.~\ref{fig:strategy_scheme}, panel 4). Then we apply an offset to place the planet onto the science fibre (Fig.~\ref{fig:strategy_scheme}, panel 5). 

The challenge to accurately inject the planet’s PSF into the science fibre relies on the high calibration accuracy performed at each procedure step. In the first strategy the critical first step is to accurately inject the stellar PSF into the centring fibre, while for the second strategy the crucial first step consists of accurately placing the stellar PSF on the intersection \I. All the following steps to inject the planet’s PSF into the science fibre will be performed blindly, relying only on the high accuracy of the calibrations computed in the laboratory and the known astrometry of the planet. In HiRISE, the science fibre \S will be connected to CRIRES+ preventing any verification of the planet's PSF injection in the science fibre during the acquisition.

In a coronagraphic configuration, the centring strategies are slightly different. Due to the presence of the occulting mask of the coronagraph, the star is not directly visible on the tracking camera. Therefore, to monitor the localisation of the star behind the mask, we use `satellite spots' artificially created by the deformable mirror in the system. These spots are routinely used in VLT/SPHERE and provide an accuracy of 1.2 mas (0.03 \lsd) for the centring of coronagraphic images \citep{vigan2016first,zurlo2016first}. The satellite spots, which are replicas of the stellar PSF, are created at a known angular separation by applying two orthogonal sinusoidal modulations on the high-order deformable mirror (HODM). In a coronagraphic context, the centring strategies presented above are modified by the necessity of first  accurately locating the stellar PSF using the satellite spots. This can be achieved by injecting each spot into the centring fibre. The diagonal intersection of these spots allows  the star's PSF position to be retrieved. Then, we place the planet on top of the centring fibre by using the calibrated interaction matrix. Finally, we apply the pre-calibrated centring-science fibre (\C-\S) offset to place the planet into the science fibre. 

In the present work, we only investigate the non-coronagraphic configuration for the three strategies. The baseline coronagraph in SPHERE is an apodised-pupil Lyot coronagraph, which is based on a pupil amplitude apodiser that causes a 50\% photon loss. The simulations of \citet{otten2021direct} have demonstrated that although the apodisation slightly improves the injection efficiency into the science fibre, the loss of photons that it induces completely outbalances this gain in the final S/N of the science signal. Moreover, to benefit from the full attenuation of the diffraction,  SPHERE should be  operated in the pupil-tracking mode \citep{beuzit2019sphere}. This enables using an optimised Lyot stop, which masks the diffraction of the telescope spider vanes, but it induces a rotation of the field of view (FoV) that would need to be compensated in the FIM. For these reasons, HiRISE will initially be operated in field-tracking without a coronagraph, so we focus on this scenario in the following sections.

\section{Fibre injection on the MITHiC testbed}
\label{sec:mithic}
\subsection{Optical setup}

\begin{figure*}
    \centering\includegraphics[width=\textwidth]{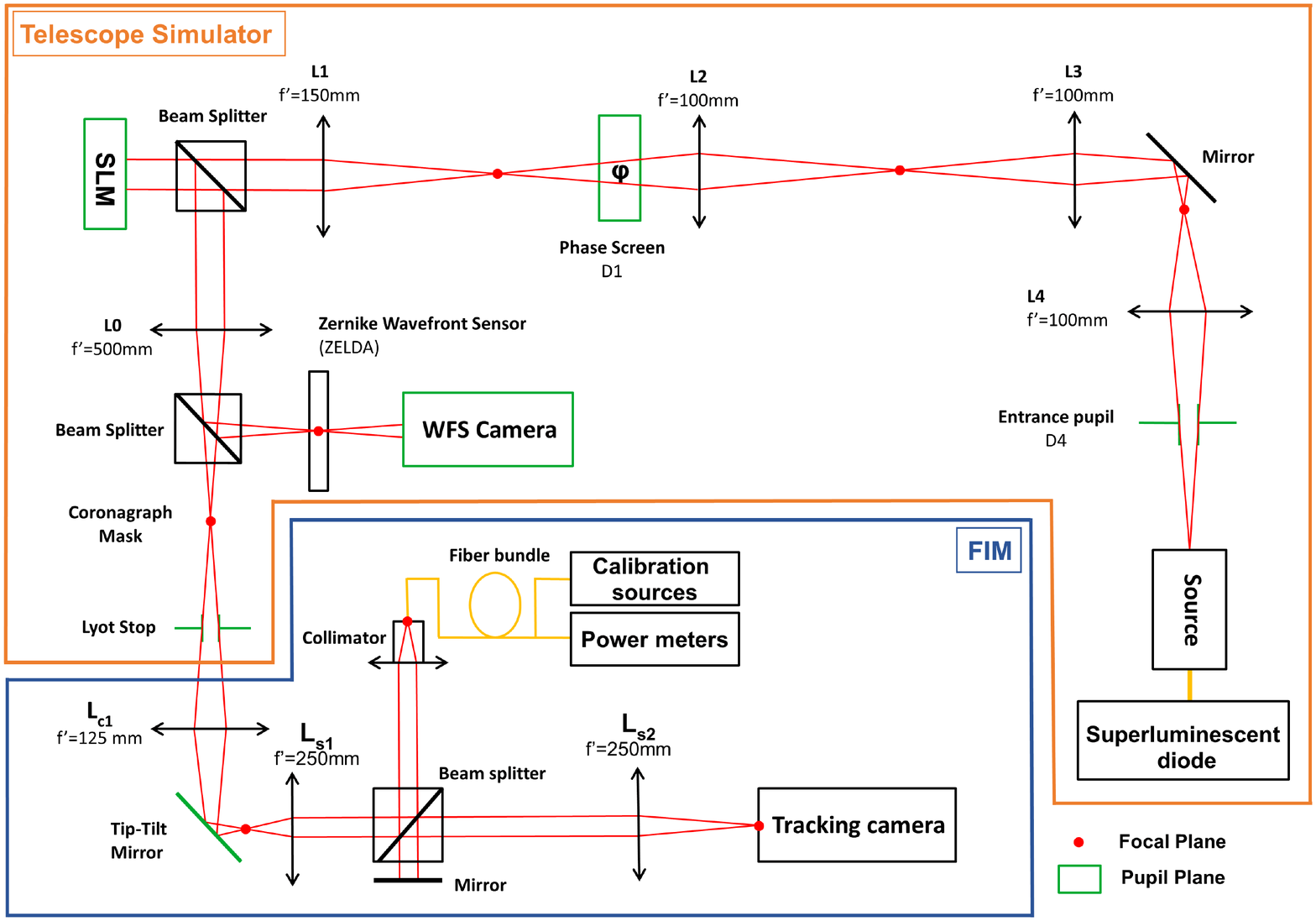}
    \caption{Schematic drawing of the MITHiC testbed. The blue box indicates the fibre injection module (FIM) and the orange box indicates the telescope simulator. Focal planes are represented as red dots and pupil planes are  in green. The bench control computer is not represented. The scale and distances between optics are not respected in this drawing.    \label{fig:mithic}}
\end{figure*}

\begin{figure}
    \centering\includegraphics[width=9cm]{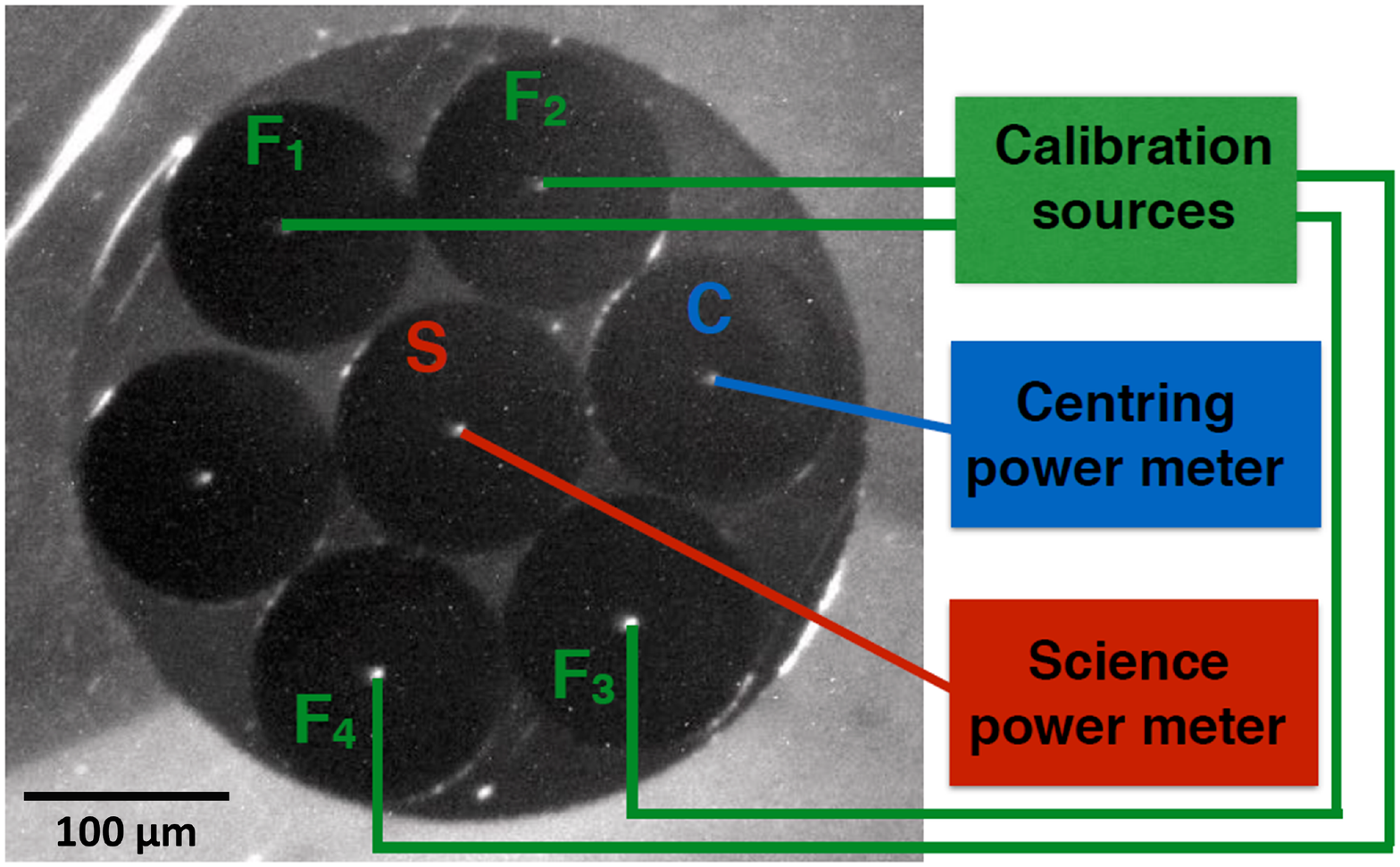}
    \caption{View of the Thorlabs fibre bundle with a binocular. The seven  dark circles are the cladding of the seven fibres (125\,\mic diameter). In the centre of each of these fibres, the core (4.5\,\mic diameter) is visible (bright spots). The science and centring fibres are respectively represented in red and blue and connected with SMF to two power meters. The four retro-injection fibres are represented in green and are connected with SMF to the retro-injection home-built setup. The spare fibre is not labelled.}
    \label{fig:FB}
\end{figure}

MITHiC is a HCI testbed located in the optics laboratory at the \emph{Laboratoire d’Astrophysique de Marseille} (France).  It was developed in 2010 and has been used to develop and test various methods for HCI, such as COFFEE \citep{paul2013high}, the ZELDA wavefront sensor \citep{n2013calibration, pourcelot2021calibration}, and the Roddier and Roddier coronagraph \citep{soummer2003a, soummer2003b,Ndiaye2010,Ndiaye2012b}. The testbed has already been described in detail in \citet{pourcelot2021calibration}, so we only mention the key elements in this section.

MITHiC is composed of a telescope simulator (TS), a wavefront sensor (WFS) to measure the static aberrations, and  a classical Lyot coronagraph (CLC) \citep{lyot1933study}, as shown in Fig.~\ref{fig:mithic}. The TS generates a PSF with a polarised monochromatic fibre-coupled super-luminescent diode at a wavelength of 670.7~nm. The spatial light modulator (SLM), which acts as a deformable mirror (DM) in the pupil plane (274 pixels across the pupil diameter), allows   a phase correction to be applied in closed-loop to flatten the wavefront and   the wavefront to be modulated for particular needs, such as creating the satellite spots or introducing known aberrations. The WFS on MITHiC is a Zernike wavefront sensor called ZELDA \citep{pourcelot2021calibration, n2013calibration}. Due to the aberrations in the MITHiC setup, if no corrections are applied on the SLM, the optical aberrations of the MITHIC bench are quantified to $\sim$33\,nm root mean square (rms), and can be decreased to 20\,nm\,rms or better with a correction based on a ZELDA measurement. 

To reproduce the setup that will be implemented in HiRISE, we modified MITHiC by adding an implementation of the FIM (Fig.~\ref{fig:mithic}). The first modification was to re-image the pupil of the bench on a mirror glued on a piezo tip-tilt mount (PI-S335) and located after the WFS stage. The mirror is oriented at 45\degre with respect to the optical axis to also fold the beam and make the whole setup more compact. Then, the beam is recollimated and split into two to implement the branches of the FIM. We use a 50/50 beam-splitter cube and not a dichroic because MITHiC is fully monochromatic. In the tracking branch, the stellar PSF is imaged using a doublet on a Coolsnap HQ2 CCD camera manufactured by \emph{Teledyne} located in a focal plane with a 4.26 pixel/(\lsd) sampling. In the rest of the paper, the Coolsnap camera will be referred to as the tracking camera. In the injection branch, the beam is refocused using a Thorlabs F810SMA-635 collimator (output $1/e^2$ beam diameter: 6.7\,\mic; wavelength: 635\,nm) to create a focal plane image at the tip of a FB.

The FB is a custom bundle manufactured by Thorlabs. It is composed of seven  4.5/125 SMF fibres, 2\,m in length, mounted with a unique SMA connector in the collimator. A picture of the tip of the bundle is provided Fig.~\ref{fig:FB}. The fibres are labelled \S for the science fibre, \C for the centring fibre, and \F{1} to \F{4} for the feedback fibres. There is one spare fibre that is not used in the setup. The seven fibres are separated from each other at the centre of the FB and the end connectors on the other side are individual FC connectors. During the alignment of the FIM, the science fibre is aligned with respect to the optical axis, while the others remain off-axis. Since MITHiC is monochromatic and does not have a dedicated spectrograph, we connect the \S and \C fibres to two different Thorlabs PM101 power meters with S150C cells (Fig.~\ref{fig:FB}, \S power meter and \C power meter) to measure the injected flux in watts. 

For the second and third centring strategy, the \F{1--4} fibres are retro-fed into the system through a home-built  system (Fig.~\ref{fig:FB}, calibration sources) based on four LEDs controlled by an Arduino micro-controller. The LEDs emit light into the SMF through the collimator, which also acts  as an injector in this configuration. The retro-fed light goes through the beam-splitter cube and is reflected to the system by a plane mirror placed closely behind the cube to avoid significant differential defocus between the tracking branch and the retro-injection. The four retro-fed fibres are finally re-imaged on the tracking camera on top of the science image.

\subsection{Limitations of the testbed}
\label{sec:limitations}

Although the MITHiC bench is within an aluminium protective enclosure, it is subject to daily variations since it is not in a temperature- and hygrometry-controlled environment. Moreover, some active components placed on the bench contribute to heating it and creating turbulence and frequency vibrations below 56\,Hz. These disturbances have been attenuated as much as possible, but some effects remain.

Several phenomena are major contributors to the limitation of the bench, such as the jitter, the drift of the PSF, and, to a much lesser extent, the quasi-static aberrations. Firstly, the jitter of the PSF is intrinsic to the turbulence or the vibration inside of the enclosure. To measure the PSF's jitter, we acquired a data of 200 frames of 1\,ms exposure time and measured the PSF's position with a 2D Gaussian fit in every frame. Based on the measurements, we estimate the standard deviation at 0.04\,\lsd for the PSF's jitter. The typical timescale of the jitter is of the order of a few milliseconds.

For the estimation of the PSF's drift, we acquired an image of the stellar PSF every 0.5\,s during 30\,min using the tracking camera. Then, the centre of the PSF was estimated in every frame with a 2D Gaussian fit. The PSF’s drift value computed from the standard deviation and is estimated for both X and Y to 0.02\,\lsd over 30\,min.

Finally, the variation of static aberrations measured with ZELDA are estimated at 7\,nm\,rms in two weeks if the bench remains in the same configuration (no opening of the panels). Since the timescale of the variations of the static aberrations is very long, this is not considered as a significant error term for our tests that take at most a few hours.

\subsection{Centring accuracy specification}

\begin{figure}
    \center
    \includegraphics[width=8cm]{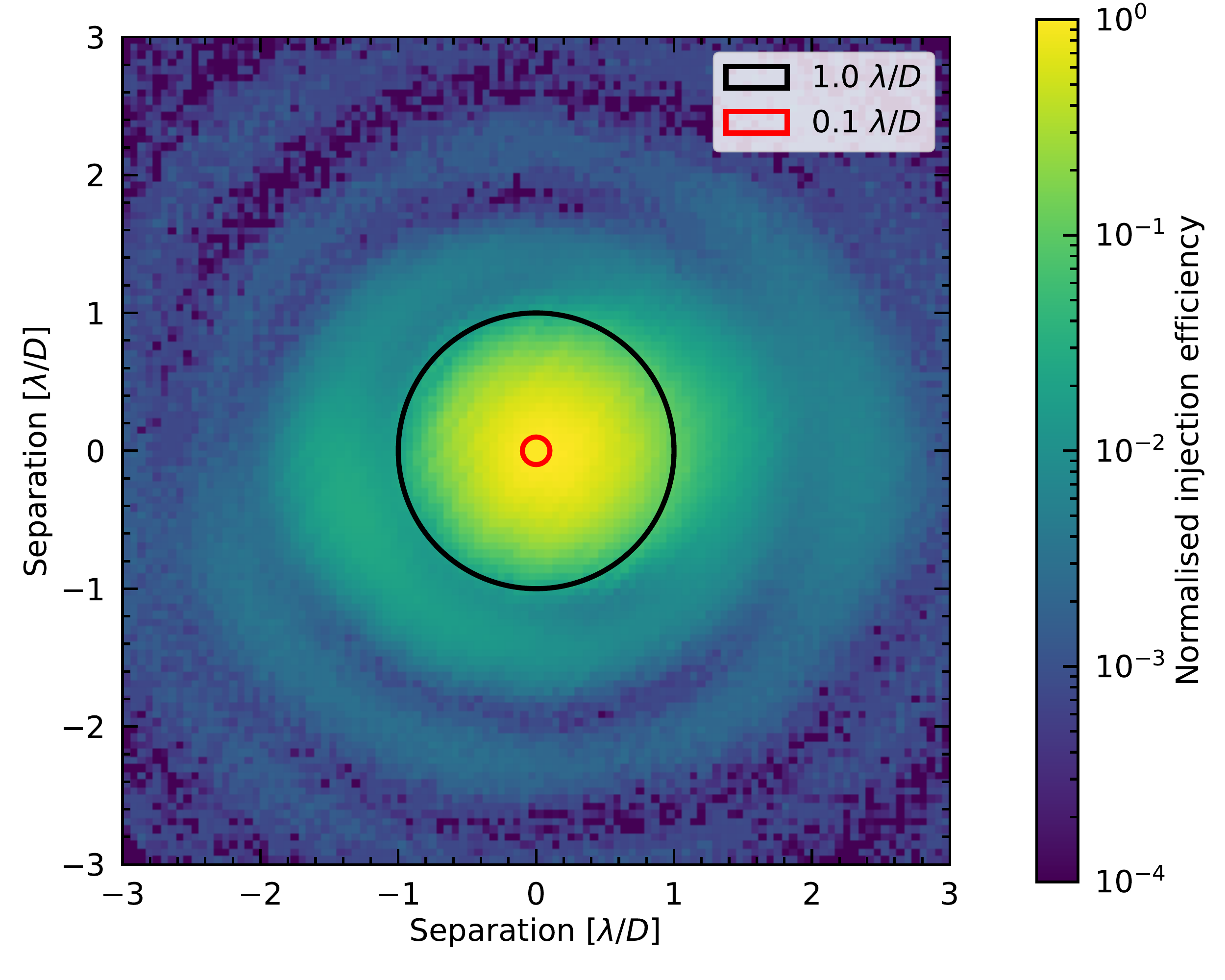}
    \caption{Injection map obtained with the centring power meter by scanning a grid of 6$\times$6\,\lsd with a pitch of 0.05\,\lsd in both directions. The red circle represents the required accuracy of 0.1\lsd.}
    \label{fig:injection_map}
\end{figure}

\begin{figure}
    \center
    \includegraphics[width=9cm]{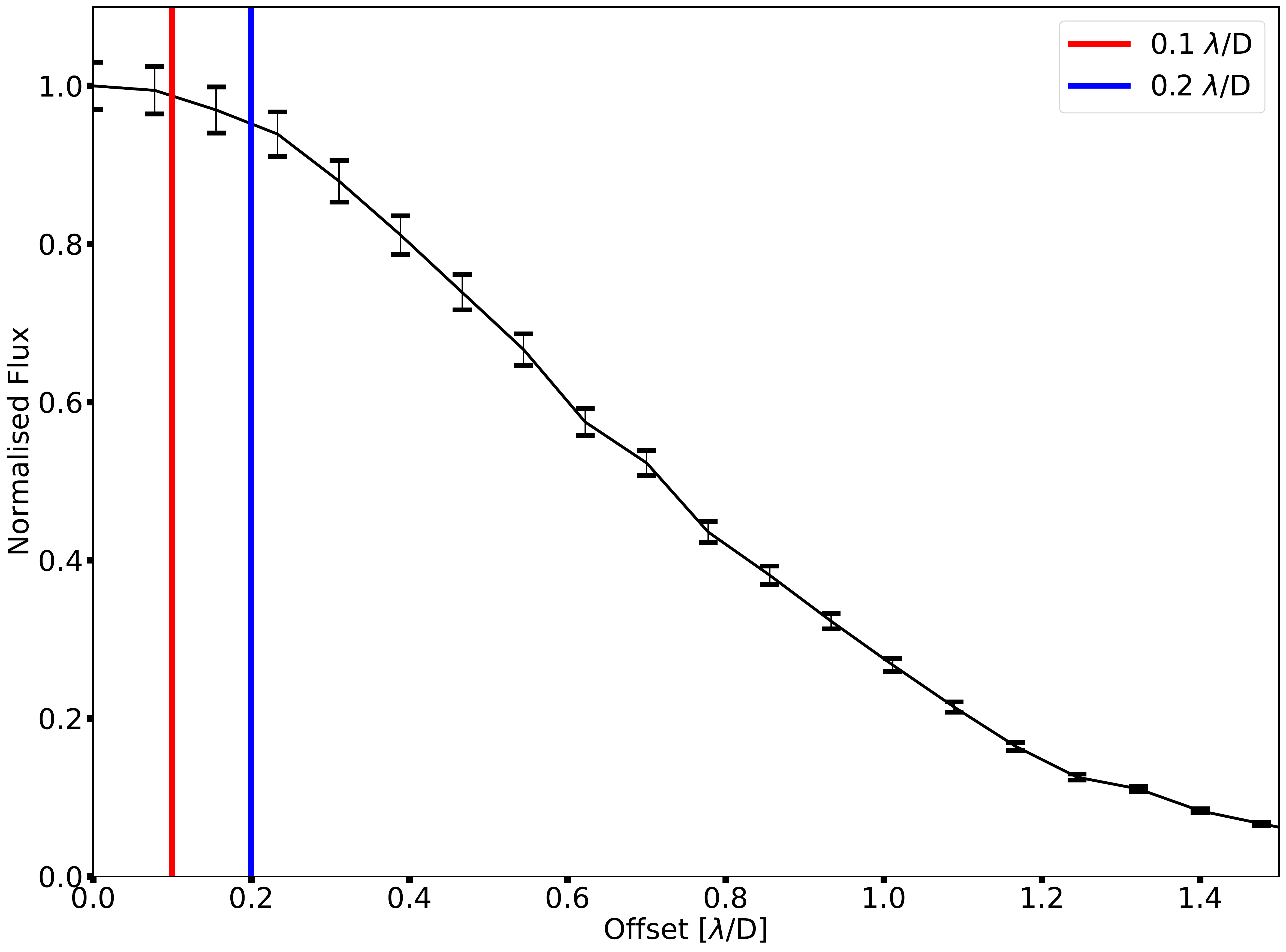}
    \caption{Radial profile of the injection map computed with the centring fibre on MITHiC.}
    \label{fig:slice_offset_psf}
\end{figure}

The performance of HDC instruments is essentially driven by the overall transmission of system. The planetary signal is several orders of magnitude fainter than the star, so the final signal can easily become limited by either photon noise (ideal case) or instrumental noise sources such as the background or readout noise (worst case). 

Injection into SMF can quickly lead to significant flux losses if the centring is not accurate, which is why this parameter is critical. The simulations by \citet{otten2021direct}  show that a centring accuracy of 0.1\,\lsd is a necessary requirement for HiRISE. With this accuracy we expect a maximum peak coupling efficiency loss of five percentage points compared to a case with  perfect centring of the PSF on the science fibre. This is a tight requirement that is illustrated in Fig.~\ref{fig:injection_map}, which shows a normalised injection map of the star's PSF into the centring fibre on MITHiC, with two circles delimiting the 0.1\,\lsd and 1\,\lsd regions around the centre. The injection map represents the flux injected into the \S and \C fibres, as measured by the corresponding power meters. 
Each point of the injection map corresponds to a tip-tilt mirror position and a PSF position in the focal plane of the tracking camera. The radial profile of the injection map shown in Fig.~\ref{fig:slice_offset_psf} demonstrates that at 0.1\,\lsd the loss is estimated to 2.6 percentage points on MITHiC. Although VLT/HiRISE and MITHiC have slightly different optical parameters, this value is consistent with the values obtained by simulation for HiRISE. A slightly lower accuracy of 0.2\,\lsd could be acceptable for HiRISE, but at the cost of a loss of approximately ten percentage points  compared to the perfectly centred case.

\section{Implementation of the centring strategies}
\label{sec:lab_centring_strat}

In this section we present the practical implementation of the centring strategies on MITHiC described in Sect.~\ref{sec:hirise} with the goal of estimating the positioning accuracy of the planet's PSF on the science fibre. We first describe some key calibrations that are required in Sect.~\ref{sec:calibrations}, then we present the data acquisition in Sect.~\ref{sec:data_acquisition}, and finally we present the performance results in Sect.~\ref{sec:results}.

\subsection{Calibrations}
\label{sec:calibrations}

\subsubsection{Centres of the fibres}
\label{sec:centres_of_fibres}

The first necessary calibration is the ability to locate the \S and \C fibres in the tip-tilt commands space, so we compute a coarse injection map. For a more accurate determination of the fibres centre, we used the tip-tilt mirror position associated with the maximum of the injection map as input to a gradient descent algorithm based on the Nelder--Meade approach, which maximises the flux injection into the fibres. Because the core of the PSF is a convex envelope, the convergence is highly repeatable and we did not encounter problems related to local minimums. From here on, this optimisation of the injection based on the gradient descent algorithm is referred to as an injection optimisation.

\subsubsection{\C-\S fibre distance}

\begin{figure}
    \center
    \includegraphics[width=0.5\textwidth]{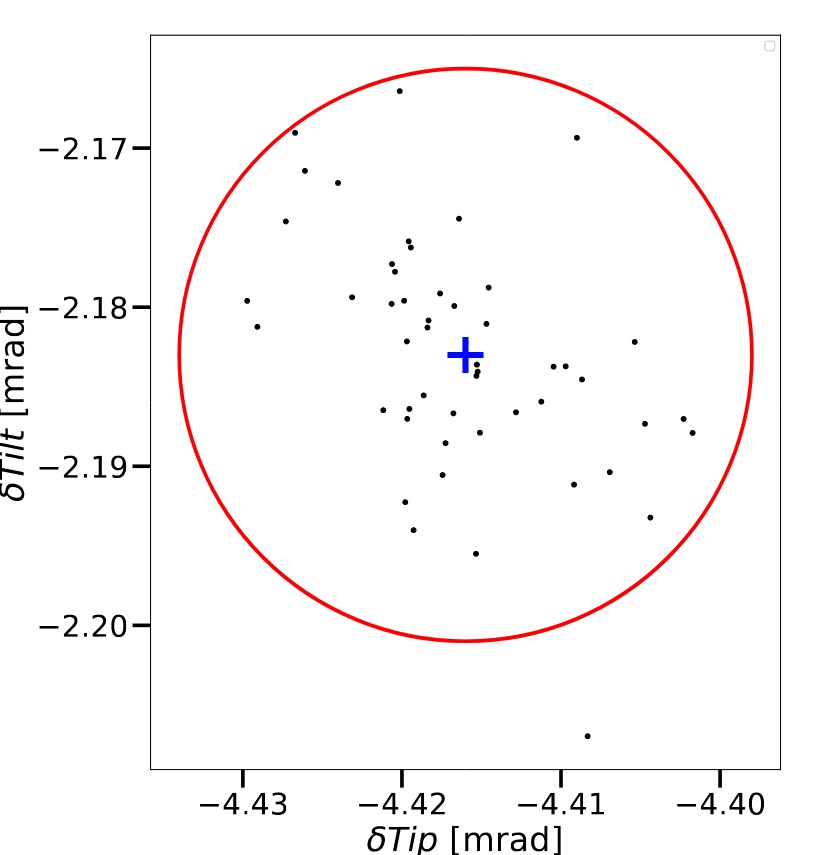}
    \caption{Measurements of the distance between the centring fibre and the science fibre for 100 consecutive tests. The red circle represents the accuracy requirement of  0.1\,\lsd. The blue cross corresponds to the mean position of -4.416 mrad in tip, -2.183 mrad in tilt and the standard deviation is estimated at 0.007 mrad in both directions.}
    \label{fig:distance_cs}
\end{figure}

The second important calibration is the accurate \C-\S distance, which is required to be able to apply a tip-tilt offset to switch between fibres in the first and third centring strategies. To calibrate this offset, we performed 100 measurements (50 measurements on each fibre) based on the following steps. First, the PSF is  centred on the \S fibre using the injection optimisation described in Sect.~\ref{sec:centres_of_fibres}; then a coarse tip-tilt offset is applied to switch to the \C fibre; and finally  the PSF is accurately centred using the injection optimisation. This provided 100 measurements of the accurate distance between the two fibres, and the final adopted value corresponds to their average. The calibration accuracy of the \C-\S distance is illustrated in Fig.~\ref{fig:distance_cs}, where each point represents the displacement position in tip-tilt before and after the optimisation. We find that 99\% of the points are within the accuracy requirement of 0.1\,\lsd with a mean estimated at -4.416\,mrad in tip and -2.183\,mrad in tilt and a standard deviation of 0.007\,mrad (0.04\,\lsd) in both tip and tilt. We conclude that the \C-\S distance calibration  matches our accuracy requirement since the standard deviation is below the requirement. Finally, we note that the standard deviation estimation is approximately equal to the PSF's jitter uncertainty, so the \C-\S distance calibration precision and accuracy are certainly limited by the jitter of the PSF.

\subsubsection{Offset between the retro-injection and tracking branches}

\begin{figure}
    \center
    \includegraphics[width=0.48\textwidth]{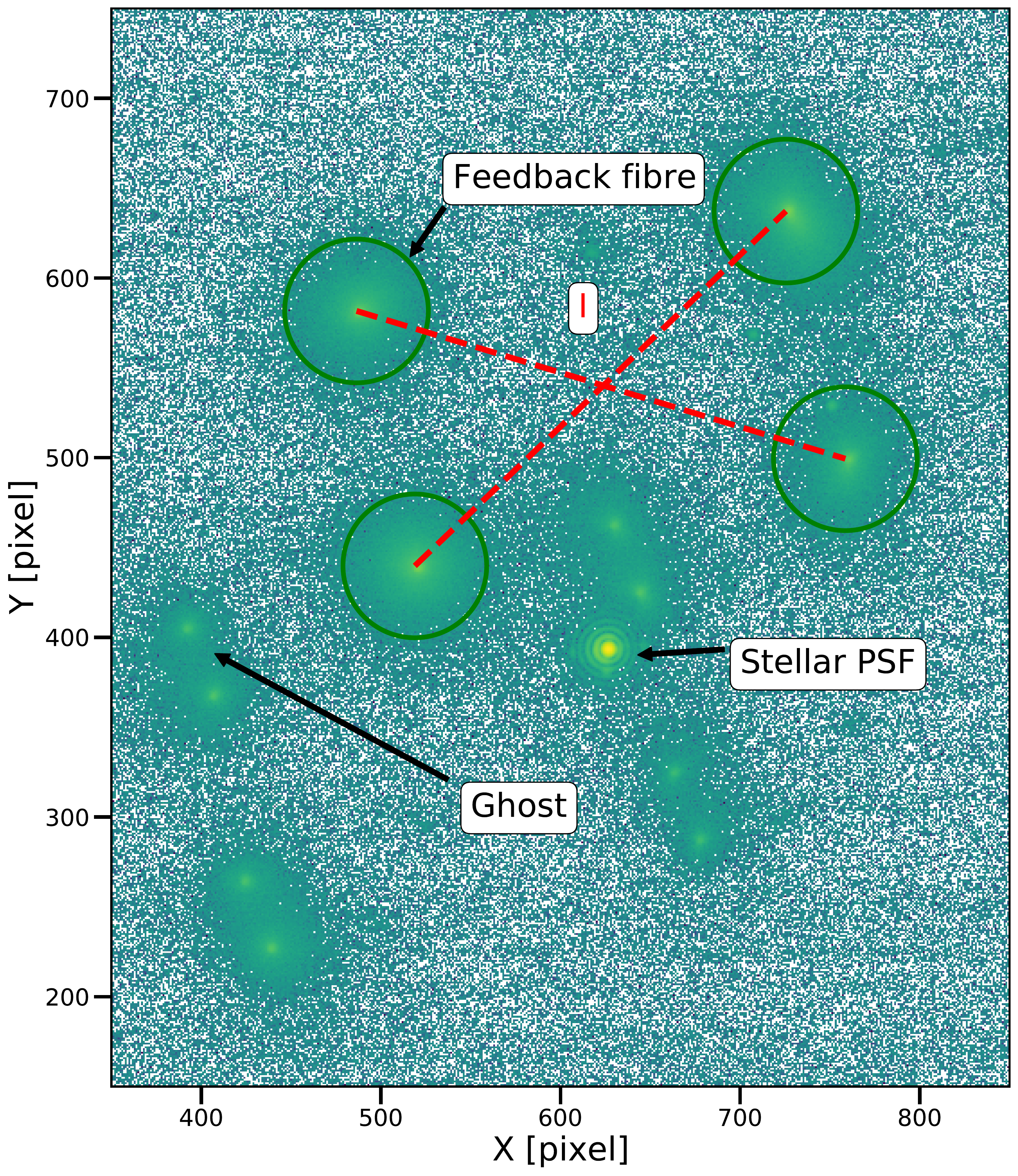}
    \caption{Log-scale image acquisition on the tracking camera. The four feedback fibres represented as green fibres are imaged in the focal plane camera and used to localise the science fibre. The beacon's spot represents the stellar PSF injected into the science SMF and the planet's PSF is not displayed. The ghosts are caused by the reflection of light in the optical system. }
    \label{fig:fov}
\end{figure}

In the second and third strategies, the calibration fibres \F{1-4} are retro-fed at the same time with a LED signal and re-imaged on the tracking camera superimposed on the science image. This is illustrated in Fig.~\ref{fig:fov}, where the stellar PSF is visible close to the centre with the four images of the retro-injection fibres. In this image the PSF was previously   centred on the science fibre using the method described above. In theory, the diagonal intersection (\I) of the four spots, retrieved by a 2D Gaussian fit, should determine the position of the science fibre on the tracking camera with respect to the retro-injection fibres. However, the mirror positioned behind the beam splitter is slightly skewed with respect to the optical axis and introduces an offset between the tracking branch and the retro-injected beam. Although the PSF is centred on the \S fibre in Fig.~\ref{fig:fov}, this is why its image on the tracking camera shows a significant offset with respect to \I. The I-S offset is accurately measured during a dedicated calibration procedure and used for the second strategy. For the third strategy, the calibration follows the same steps, but measures the \I-\C offset.

\subsubsection{Accurate positioning of the PSF}
\label{sec:accurate_positioning_psf}

\begin{figure}
    \center
    \includegraphics[width=0.48\textwidth]{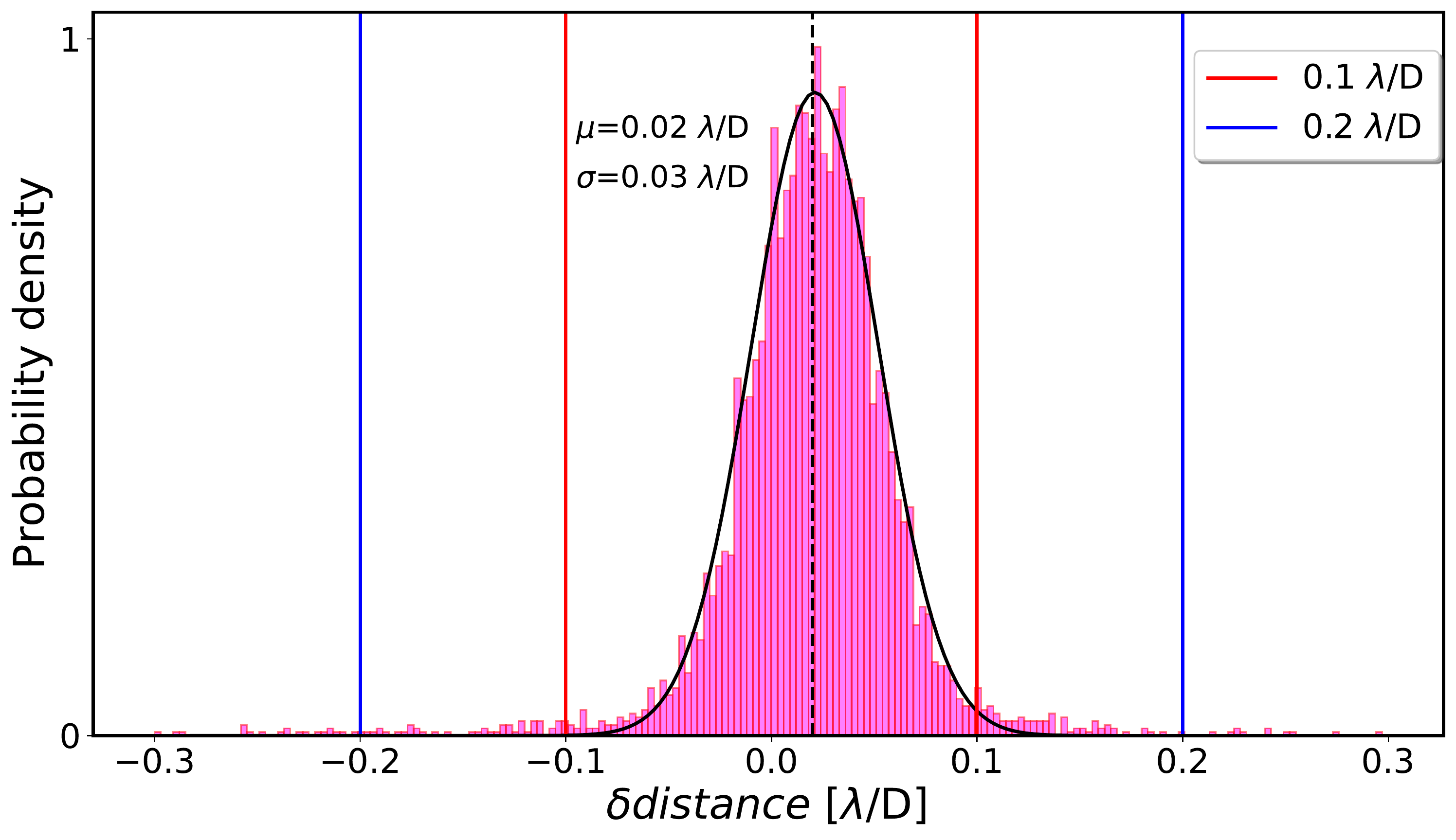}
    \caption{Distribution of the residual interpolation function: the probability density as a function of the $\delta$ distance. It represents the accuracy of the interpolation function to place the stellar PSF at any position on the camera FoV.}
    \label{fig:histo}
\end{figure}

\begin{figure}
    \center
    \includegraphics[width=9cm]{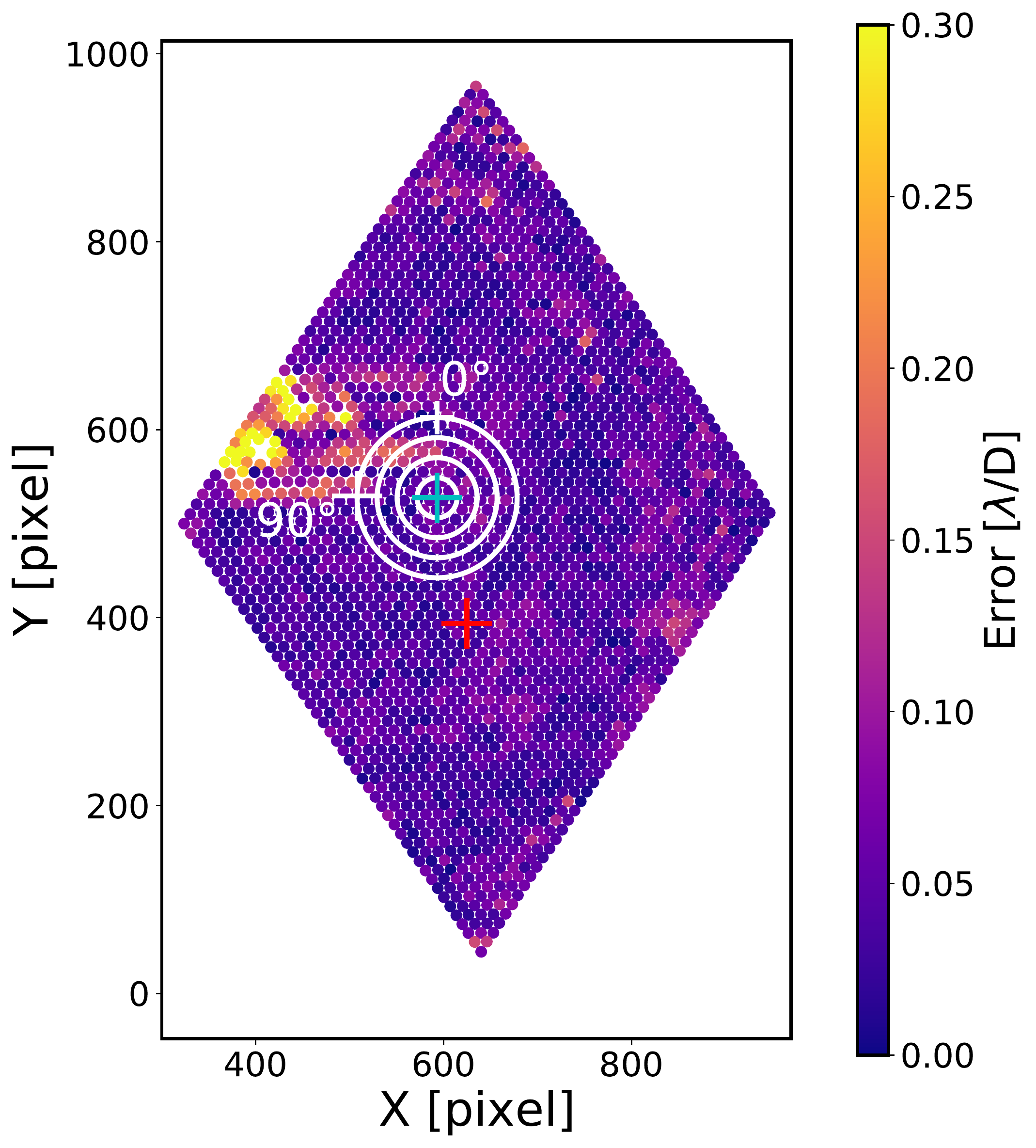}
    \caption{Residual of the interpolation function as measured on the tracking camera. The red cross and the blue cross  give the positions of the science and centring fibre, respectively. The concentric rings in white represent the different angular separations (5, 10, 15 and 20 \lsd) and the different angle positions of the planet when the star is well centred on the centring fibre.
    The colour bar corresponds to the uncertainty error on the interpolation function. We note that  97.61\% of the points mentioned in Fig.~\ref{fig:histo} are within the specification accuracy of 0.1\,\lsd.}
    \label{fig:map_uncertainty}
\end{figure}

A final  calibration is required to be able to accurately position the stellar PSF at any coordinate on the tracking camera. We use a non-linear approach with an interpolation function based on a full 2D calibration of the response of the tip-tilt mirror \citep{yelda2010improving,service2016}, as seen by the tracking camera. To create the interpolation function, we build a calibration grid by scanning a 12.5$\times$12.5\,mrad area in tip-tilt, with a pitch of 2\,mrad. For each tip-tilt position, the stellar PSF is imaged on the tracking camera and a 2D Gaussian fit is used to compute the PSF's centre in pixels. The Gaussian fit accuracy has been estimated to $\sim$0.01\,pix, which corresponds to 0.03\,\lsd. The calibration grid is then provided as  input data to the \texttt{LinearNDInterpolator} function from the \texttt{scipy} library \citep{2020SciPy-NMeth}, which uses the Delaunay triangulation method to perform interpolations in N dimensions. Using the resulting interpolation function, it is possible to obtain the tip-tilt command to place the stellar PSF at any position in the camera's FoV.

An alternative approach would be to use a classical AO-like interaction matrix, where the relationship between the focal plane position and the tip-tilt commands is calibrated by introducing a single known offset in tip and tilt. However, this approach assumes that the offset value does not matter because there is a strictly linear relationship. This is not the case in our system because of optical distortions introduced between the tip-tilt mirror and the tracking camera, which is why an interpolation grid was adopted.

We estimated the accuracy of the positioning by using the interpolation function to place the PSF on a test grid of 48$\times$48 points interleaved with the original calibration grid. For each target position, the position of the PSF was estimated on the tracking camera image and compared to the target position to compute the residuals. We find that  6\% of the outliers  are located at more than 20$\sigma$ from the sample data mean. We reject all outliers located beyond three standard deviations from the sample data mean. The histogram of the residuals for the full test grid is displayed in Fig.~\ref{fig:histo}. The residuals are quasi-Gaussian with a standard deviation of 0.03\,\lsd, which is well below 0.1\,\lsd. Even taking into account the wings of the distribution, 97.61\% of the residuals fall within $\pm$0.1\,\lsd.

Even though the accuracy for most positions is well within our specification of 0.1\,\lsd, the presence of outliers can potentially have a negative impact. In Fig.~\ref{fig:map_uncertainty} we plot the measured position on the tracking camera of all points in the test grid, colour-coded with the position error with respect to the requested target position. The positions of the \C and \S fibres are indicated with crosses. We note an area of particularly poor accuracy located close to the position of the \C fibre, in the direction that corresponds to the  planet's position angles (in tip-tilt space) between $\sim$45\degre and $\sim$100\degre. For those tip-tilt commands, the beam starts hitting the edge of one of the lenses, which impacts significantly the PSF’s image quality on the camera. A combination of aberrations is noticed in the PSF’s shape,  the coma being the main contributor, which affects significantly the shape of the PSF. As a result, the 2D Gaussian fit of the PSF is less accurate, which is problematic since the PSF’s centroid is used to create the interpolation calibration grid. The presence of the outliers could explain the loss of accuracy.

The calibration grid used to build the interpolation function is based on the acquisition of the stellar PSF on the tracking camera; consequently, its stability relies on the bench limitation. The calibration grid takes approximately 6 minutes to be computed and is stable over approximately 30 minutes in optimal conditions due to the PSF's drift estimated in Sect.~\ref{sec:limitations} at 0.02\,\lsd.

\subsubsection{ZELDA calibration}

The overall optical quality of the bench would normally allow  the centring strategies to be investigated without requiring a dedicated optimisation of the wavefront. However, the standard operations of the MITHiC testbed include a calibration of the aberrations of the bench using the ZELDA wavefront sensor \citep{pourcelot2021calibration}. We maintain this calibration at the beginning of each test as a sanity check and to benefit from the best possible PSF quality. A wavefront quality better than 20\,nm\,rms is typically obtained.

\subsection{Data acquisition}
\label{sec:data_acquisition}

In this section we test the centring strategies and quantify their respective accuracy. We recall that the goal, in the context of HiRISE, is to place the planet's PSF as close as possible to the centre of the \S fibre so as to maximise the injection efficiency and therefore the planetary flux that will be transmitted to a high-resolution spectrograph.

On MITHiC we do not have the capability to create a true off-axis point source that would simulate a planet. To create the planet's PSF, we introduce a sinusoidal modulation on the SLM that creates two symmetric satellite spots and take one of these spots as the planet's PSF. The spatial frequency and orientation of the sine wave enables choosing the angular separation and position angle of the planet. Contrary to an actual planet for which the astrometry will be known beforehand based on previous direct imaging measurements, in our tests we measure the relative star-planet astrometry on a tracking camera image just after the planet has been introduced with the SLM. The amplitude of the sine wave is set to 100\,nm\,rms, which results in a star--planet contrast ratio of a factor 3.48. The contrast is modest, but does not impact the final performance results because on-sky the system is  blind to the planet's signal during the centring procedure. However, having a modest contrast ratio in the laboratory is necessary to check the final accuracy.

At the end of each centring procedure (regardless of the strategy), a verification optimisation (VO) is performed to evaluate how close the planet's PSF is to the centre of the \S fibre. The displacement in tip and tilt between the position before the VO and after the VO, $\epsilon$, provides a direct measurement of the distance to the centre. This $\epsilon$ is the error that defines the success or failure status of the performance test by comparing it to the specification accuracy of 0.1\,\lsd. However, the VO is performed only in the laboratory and cannot be used on-sky to test if the planet is accurately injected into the science fibre. The science fibre will be connected to the CRIRES+ spectrograph and in any case the planet's signal will be several orders of magnitude fainter than that of the star.

For the performance study, we performed five series of 20 centring tests for planets introduced at 12 different position angles around the star. The star--planet angular separation was set to 10\,\lsd for all tests, which corresponds to the average separation where known companions have recently been imaged with HCI. We had to take into consideration the drift of the PSF (see Sect.~\ref{sec:limitations}) and its impact on the temporal stability of the calibration grid. This is why between each series we recompute the calibration grid used as   input to the interpolation function. For each position angle, all the $\epsilon$ values are combined to compute an overall success rate when compared to centring accuracy specifications of 0.1, 0.2, and 0.3\,\lsd. This repeatability test is decisive to achieve the performance study since it highlights the accuracy of each strategy. 

\subsubsection{Strategy 1}

The centring procedure starts by measuring an injection map to coarsely position the star's PSF on the \C fibre, then an injection optimisation is performed to accurately centre the star on the \C fibre. The pre-calibrated \C-\S offset is applied to the tip-tilt command to position the star on the science fibre as an intermediate step. Then we use the pre-calibrated interpolation function to measure the offset command to switch between the stellar PSF and the planetary PSF on the science fibre, based on the relative astrometry computed on the tracking camera images. Finally, a VO is performed to evaluate the $\epsilon$ error.

\subsubsection{Strategy 2}

The second strategy starts by switching on the four feedback fibres and taking an image on the tracking camera. In the focal plane the position of the four feedback fibres is determined by a 2D Gaussian fit. The positions are used to compute the intersection \I and compare its position to the stellar PSF. Then, by applying a tip-tilt command retrieved by the interpolation function, the stellar PSF is placed into the science fibre by using the previous \I-\S calibration. The offset command in tip-tilt to move the planet into the science fibre is calculated using the same method described previously. Finally, the VO is performed to estimate the final $\epsilon$ error.

\subsubsection{Strategy 3}

The third strategy combines the first and the second strategies. The initial step is to scan the mirror in tip-tilt to coarsely position the star on the \C fibre and perform an injection optimisation to refine the centring. Then, we use the same steps described for the second strategy to place the star into the centring fibre based on its relative position with respect to \I. The calibration is now not performed between \I and \S but between \I and \C. After the star is positioned on the \C fibre, we apply the corresponding tip-tilt offset command to place the planet on the \C fibre and we apply the pre-calibrated offset to switch from the centring fibre to the science fibre. Finally, a VO is performed to check if the planet is accurately injected into the science fibre and to measure the error $\epsilon$.

\subsection{Results}
\label{sec:results}

\begin{figure}
    \center
    \includegraphics[width=9cm]{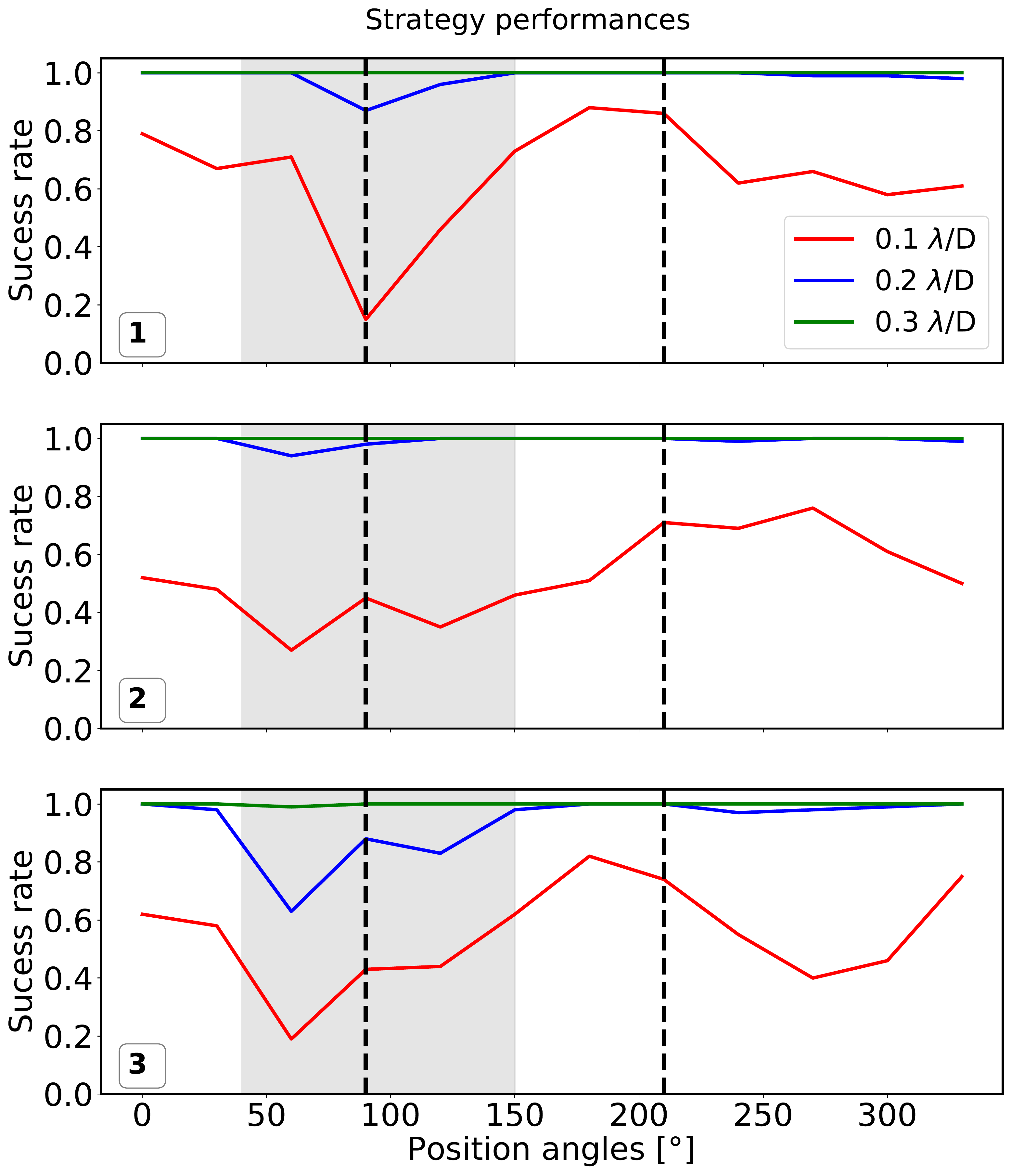}
    \caption{Success rate as a function of position angle around the star. The results for the first, second, and third strategies are plotted in the upper, middle, and lower panels, respectively. The grey area represents the region where   an emphasised trend is found. The black dashed lines at 90\degre and 210\degre represent the angles for which additional tests are performed and demonstrated in Fig.~\ref{fig:results_hypothesis}.}
    \label{fig:results}
\end{figure}

\begin{figure*}
    \center
    \includegraphics[width=\textwidth]{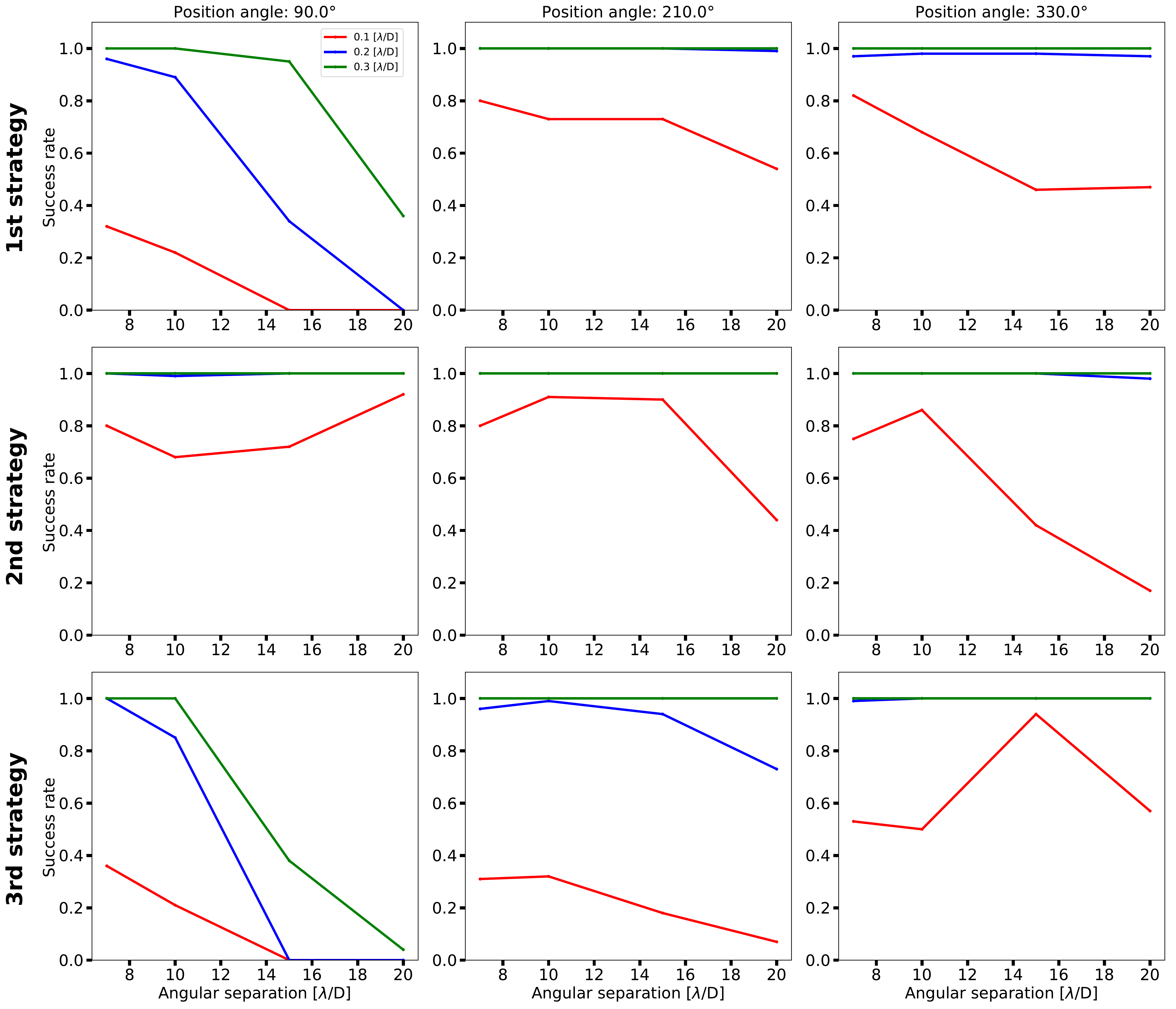}
    \caption{Success rate as a function of the angular separation from the star. The results for the first, second, and third strategies are plotted in the upper, middle, and lower panels, respectively. The black dashed lines at 90\degre and 210\degre represent the angles for which additional tests are performed and demonstrated in Fig.~\ref{fig:results_hypothesis}.}
    \label{fig:results_lod}
\end{figure*}

\subsubsection{Performance with position angle}

Our results for a planet located at different position angles around the star and with an angular separation of 10\,\lsd are summarised in Fig.~\ref{fig:results}. The figure shows the success rate of each strategy for three specification accuracies of 0.1, 0.2, and 0.3\,\lsd.

We note a correlated trend between all strategies, particularly emphasised in the plot’s 40--150\degre region (Fig.~\ref{fig:results}, grey region). In this region all the strategies show a clear decrease in the success rate. This region roughly corresponds to the range of angles where a significant number of outliers have been identified in the calibration of the interpolation function (see Sect.~\ref{sec:accurate_positioning_psf}). This is very likely correlated to the poor results now obtained in this region. Since this is an issue related to our specific setup, we do not consider the consider the results in the 40--150\degre range as representative of the final expected performance.

For the first strategy (Fig.~\ref{fig:results}, top panel), we never reach a success rate of 1.0 for a specification accuracy of 0.1\,\lsd or 0.2\,\lsd for all position angles, while a success rate of 1.0 is reached for all position angles when considering a specification accuracy of 0.3\,\lsd. For a specification of 0.1\,\lsd the success rate goes as low as 0.2, but is on average around 0.7. For a specification of 0.2\,\lsd the success rate is significantly improved, with a value of 1.0 for most position angles and a drop at 0.8 for only a single point in our measurements.

For the second strategy (Fig.~\ref{fig:results}, middle panel), the success rate for a specification of 0.1\,\lsd never falls below 0.3, but is on average slightly lower than for the first strategy ($\sim$0.6). The success rates for specification accuracies of 0.2 and 0.3\,\lsd are above 0.95 for all position angles.

Finally, the results of the third strategy (Fig.~\ref{fig:results}, bottom panel) show a more or less equal performance as the first strategy for a specification of 0.1\,\lsd, but the results for a specification of 0.2\,\lsd appear worse than for the other two strategies. For a specification of 0.1\,\lsd, the average success rate is around 0.5, with less extreme variations inside and outside of the grey  area. However, the success rate values are clearly lower in the grey area for this third strategy. Since the third strategy is a combination of the first and second strategies, the fact that we do not find a significant increase of the success rate for the 0.1\,\lsd specification is not surprising.

\subsubsection{Performance with angular separation}

We also tested the three strategies with planets located at different angular separations. The results are presented in Fig.~\ref{fig:results_lod} for three selected position angles: 90\degre, 210\degre, and 330\degre. The 90\degree{} position was selected because it is at the centre of the grey area in Fig.~\ref{fig:results}, where the performance is particularly low. The other two position angles are separated by 120\degree. We first note that the success rate obtained at a separation of 10\,\lsd is consistent with the previous test. However, there are  minor differences that can be explained by the fact that the results shown here were acquired in a completely separate data acquisition sequence.

For the first strategy (Fig.~\ref{fig:results_lod}, top panel) the angle 90\degre shows the lowest success rate since it drops drastically as a function of the angular separation. It reaches 0.0 at worst with the specification of 0.1\,\lsd and 0.2\,\lsd, and it drops below 0.4 for the 0.3\,\lsd specifications for planets at 20\,\lsd. In contrast, for the two other position angles, the success rate is higher than 0.4 at all separations for the 0.1\,\lsd specification, and better than 0.95 at all angles for 0.2\,\lsd and 0.3\,\lsd.

The second strategy (Fig.~\ref{fig:results_lod}, middle panel) shows a success rate higher than 0.9 with an accuracy of  0.2\,\lsd and 0.3\,\lsd at all position angles and all angular separations. For a specification of 0.1\,\lsd the success rate is on average below 0.8 and shows a decrease with angular separation. There is no apparent strong correlation with the position angle for the 0.1\,\lsd specification.

For the third strategy (Fig.~\ref{fig:results_lod}, bottom panel), the results are consistent with the ones previously seen in Fig.~\ref{fig:results}: the success rates for the different specifications at position angles of 90\degre and 210\degre are worse than for the first or second strategy, and at 330\degre they are similar. The drop in performance for the three specifications at a position angle of 90\degre is even sharper than for the first strategy, confirming that whatever limits the performance of the first strategy has an even stronger impact in the case of the third strategy.

\subsubsection{Understanding the limitations}

\begin{figure}
    \center
    \includegraphics[width=6cm]{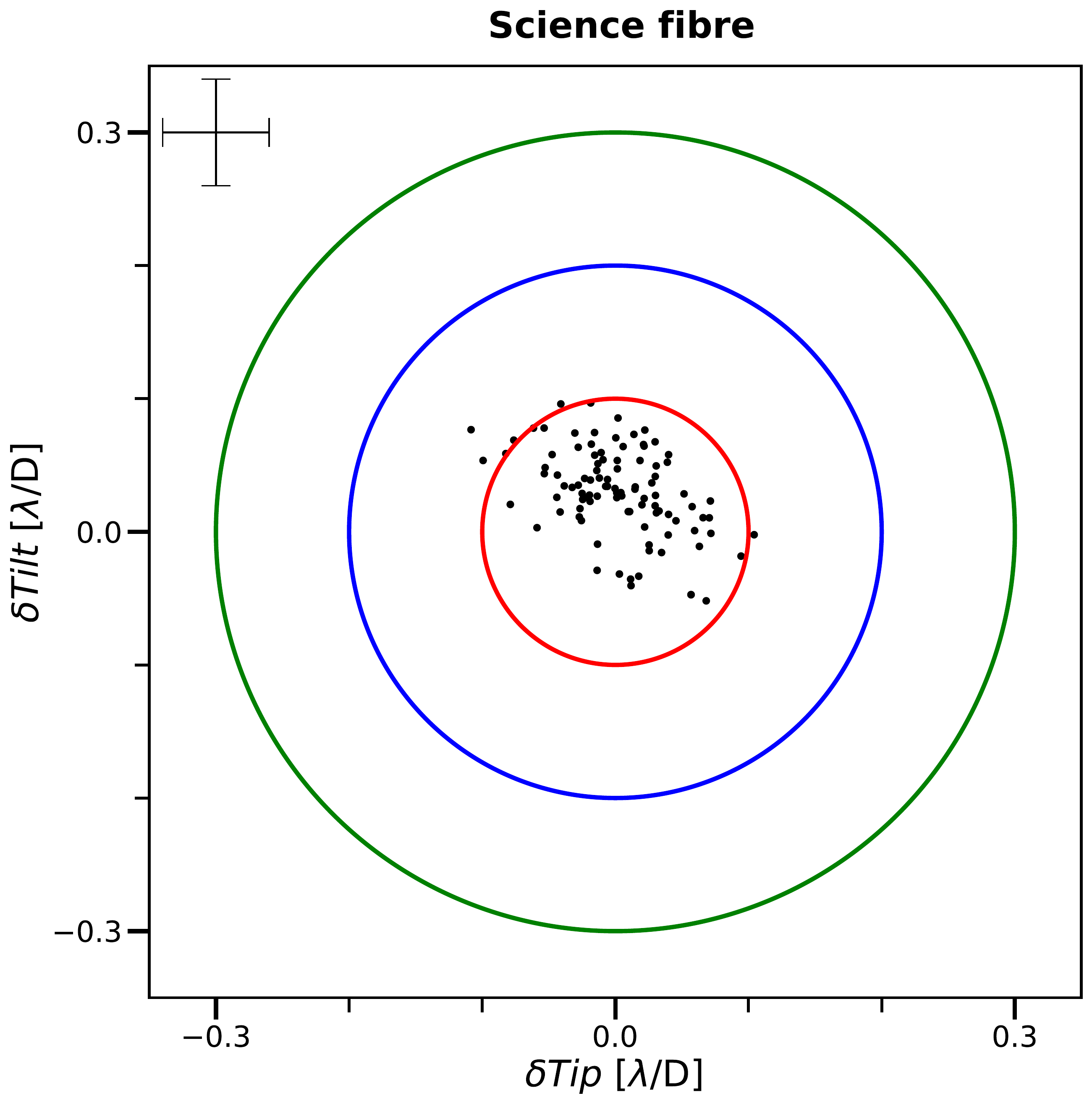}
    \caption{Success rate in the science fibre for the positioning of the star. Each dot is obtained for one entire centring procedure. The circles correspond to the different accuracies and the error bar gives   the PSF's jitter uncertainty.}
    \label{fig:star_c_s}
\end{figure}

\begin{figure}
    \center
    \includegraphics[width=9cm]{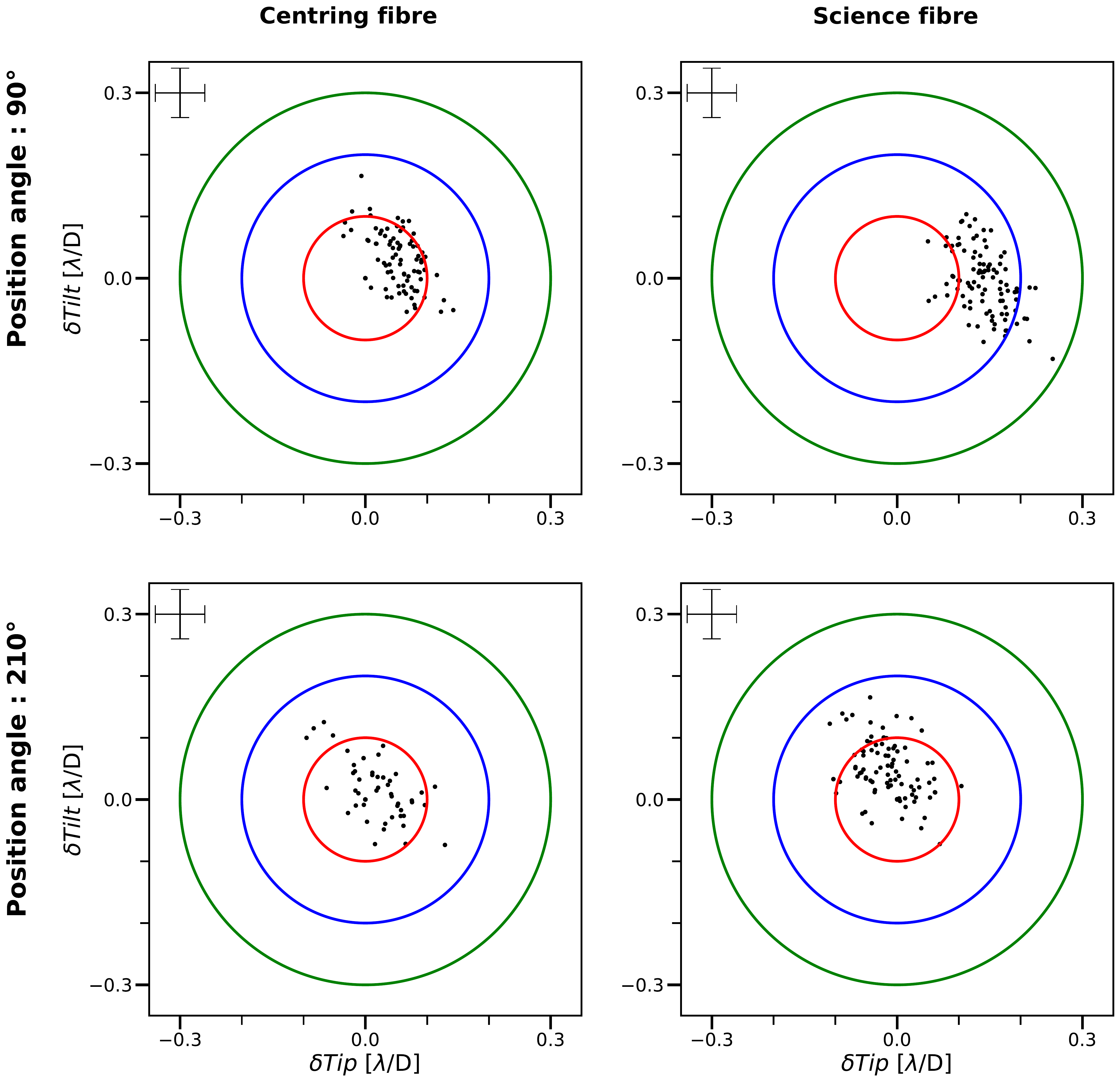}
    \caption{Success rate in the centring and science fibre for the positioning of the planet. Top panel: Success rate in the centring and science fibre for the planet at a position angle of 90\degre from the star. Bottom panel: Same results, but for a position angle of 210\degre. Each dot is obtained for one entire centring procedure. The circles correspond to the different accuracies and the error bar gives for the PSF's jitter uncertainty.}
    \label{fig:results_hypothesis}
\end{figure}

The results above demonstrate that reaching a specification accuracy of 0.1\,\lsd is extremely challenging, whatever the centring strategy. In the case of the first and third strategies we hypothesise that the loss of accuracy in the centring of the planet likely occurs at the level of the switch between the \C fibre and the \S fibre, although the \C-\S distance is accurately pre-calibrated using the tip-tilt mirror to inject the light,  alternating between the two fibres (see Sect.~\ref{sec:calibrations}).

To validate this assumption, we conduct a series of additional tests. In the first we follow the process of the first centring strategy with the goal of simply centring the stellar PSF on the \S fibre. The results for 100 independent tests are presented in Fig.~\ref{fig:star_c_s} and show a success rate of 0.94 for a 0.1\,\lsd specification and 1.0 for a specification of 0.2\,\lsd on the \S fibre. We see that on the science fibre most points are well centred, although there is a visible scatter. Nonetheless, the success rate is very high and the test demonstrates that we are able to accurately place the stellar PSF on the science fibre.

The second test is similar, but this time it is performed on planets located at a separation of 10\,\lsd and position angles of 90\degre and 210\degre. The results for 100 independent tests are shown in Fig.~\ref{fig:results_hypothesis}. At both position angles the centring is both relatively precise and accurate on the centring fibre, but the accuracy is lost when switching on the science fibre, especially for the 90\degre position angle. At this position angle, the success rate is almost zero with the 0.1\,\lsd specification and >0.9 with the 0.2\,\lsd specification. The PSF's jitter uncertainty appears compatible with the scatter of the points for the 100 tests; however, it cannot explain the loss of accuracy so there must be another limitation in the system.

The interpolation function, which is used in all three strategies to place the stellar PSF at any position in the FoV, is potentially the limiting factor. We show  in Sect.~\ref{sec:accurate_positioning_psf} that a significant part of the tip-tilt space close to the centring fibre \C in the range of position angles 45--110\degre shows a reduced accuracy that is  likely due to optical aberrations on the testbed. Interestingly, this range closely corresponds to the range of position angles in Fig.~\ref{fig:results} where the success rates significantly decrease. Figure~\ref{fig:map_uncertainty} also shows that the accuracy is worse when going away from the \C fibre's position (i.e. for planets injected at larger angular separations). This is highly consistent with the drop in performance observed with angular separation in Fig.~\ref{fig:results_lod}. We conclude that for planets in the 45--110\degre range of position angles, the centring on the \C fibre is biased because of the limited accuracy of the interpolation function and the error gets amplified when switching to the \S fibre.

For the second strategy, the \C-\S offset is not involved since we use the \I-\S distance to place the planet’s PSF into the science fibre. However, the low success rate as a function of the angle positions for a specification accuracy of 0.1\,\lsd could be explained by the defocus of the four retro-injected spots imaged on the camera. It impacts the accuracy of the 2D Gaussian fit and consequently the accuracy of the \I position determined on the camera. The loss of accuracy with angular separation is actually less pronounced than for the other two strategies, so it could be that several error factors add up and limit our accuracy to only 0.2\,\lsd.

\section{Conclusions and discussion}
\label{sec:ccl}

The first exoplanet was detected a few decades ago, and thousands of them have been detected and confirmed since then by indirect or direct techniques. The characterisation of these exoplanets remains crucial to understanding their atmospheric composition, and their formation and evolution mechanisms. The combination of HCI with HRS promises great potential to answer some of these fundamental questions. 

HiRISE proposes to implement a coupling between SPHERE and CRIRES+ at the VLT. One of the main constraints in the project is the ability to inject a known planet's PSF on the science fibre to transmit it to the spectrograph with the highest possible throughput. Previous studies for HiRISE have determined that a specification of 0.1\,\lsd would be the best choice to maximise the transmission \citep{otten2021direct}.

To study the best centring approach, we used the MITHiC testbed at LAM, which has been upgraded to closely emulate the setup used in HiRISE. The FIM part is divided into three branches: the science branch that images the focal plane on the tracking camera, the injection branch that injects the light into a small bundle of fibres using a collimator, and the retro-injecting branch following the reverse injection branch and images feedback fibres on the tracking camera. With this setup, three strategies were   implemented and tested on MITHiC to assess the feasibility of reaching a specification of 0.1\,\lsd. 

We conducted a performance study to verify the planet’s PSF centring on the science fibre \S for several angular separations and for several position angles around the star. Regardless of the strategy, we conclude that reaching an accuracy of 0.1\,\lsd is extremely challenging. It requires a high level of accuracy at every calibration step, which is difficult to reach, especially in a non-optimised system built from off-the-shelf components. In our case the final performance over a range of  planet position angles (or conversely a range of tip-tilt positions) can be explained by a lower accuracy in the calibrations used to performed the first and third centring strategies.

The first performance tests were achieved for several planet position angles with a separation angle of 10\,\lsd. For the first strategy we reach an average  success rate of 0.7 for the specification accuracy of 0.1\,\lsd, and 0.95 or higher for accuracies greater than   0.2\,\lsd. For the second strategy the success rate is on average equal to 0.6 for an accuracy of 0.1\,\lsd, and again the success rates are higher than 0.95 for an accuracy of 0.2\,\lsd and 0.3\,\lsd. Finally, the third strategy demonstrates a success rate slightly worse than the first and second strategies.

The second performance tests were performed with planets located at different angular separations; we only tested three position angles: 90\degre\ (located in the grey area), 210\degre, and 330\degre. The first strategy shows the lowest performance for 90\degre, with a success rate that severely drops as a function of the angular separations. It reaches 0.0 at worst for an accuracy of 0.1\,\lsd and 0.2\,\lsd, and is below 0.4 for an accuracy of 0.3\,\lsd. This is not surprising as this position angle is clearly identified as problematic in the calibrations. For the two other position angles we reach a success rate higher than 0.4 for 0.1\,\lsd and higher than 0.95 for an accuracy of 0.2\,\lsd and 0.3\,\lsd. The observed trends for the first strategy are found again in the results for the third strategy, but with even poorer results. Finally, the results for the second strategy are much better, with very high rates at all position angles for the 0.2 and 0.3\,\lsd specifications, and slightly decreased success rates for a specification of 0.1\,\lsd, but not as bad as for the first and third strategies.

Some aspects related to performance can be explained by limitations in the system, such as the PSF jitter and the accuracy of the interpolation function. The interpolation function is used in all three strategies to place the stellar PSF at any position in the FoV. Even though most of the interpolation function's uncertainty distribution is within the accuracy specification of $\pm$0.1\,\lsd, the presence of outliers  negatively impacts the interpolation function in certain areas. These outliers originate from hardware limitations inherent to the MITHiC testbed, which hopefully will not be present for HiRISE.

The FIM implemented in SPHERE will benefit from the complete instrument's infrastructure (stability, high-quality optics, ExAO, coronagraph). As in the MITHiC testbed implementation, we will have an optimised system with a dedicated \C fibre connected to a power meter in HiRISE, which will allow  the stellar PSF injection to be verified during the centring procedure. However, a major difference between the laboratory environment and the final implementation is that the \S fibre will be connected to the CRIRES+ spectrograph. This will prevent the output flux from being measured directly, which will affect the centring of the stellar or planetary PSF on the science fibre.

From the operational point of view there are important differences between the three centring strategies. The first strategy mainly relies on an accurate determination of the \C-\S distance, which will have been calibrated in the laboratory during the assembly integration test phase (AIT). The main advantage of this strategy is that no flux measurement is theoretically necessary at the output of the bundle on a daily basis. On the other hand, we see on MITHiC that issues related to the differential distortion between the injection and science branches, or to the accurate calibration of the interpolation matrix, can significantly impact the accuracy when injecting an off-axis point into the fibre instead of the star. At this stage it is still uncertain if the same problems will arise on HiRISE, where we have an optimised system. Fortunately, SPHERE includes a distortion calibration grid \citep{Wildi2010} that could be extremely useful to fully characterise the differential distortion in the final system.

The second strategy is inspired by the first calibration step used on KPIC phase I \citep{2021delorme}, so it has the advantage of already being demonstrated on-sky. They report that they are able to place the planet's PSF on the science fibre with a precision of less than a 0.2\,\lsd ($<$10 mas) in $K$ band, which is similar to the values that we obtained on MITHiC. For HiRISE, the retro-injection takes into account the differential distortion between the two branches. On the other hand, it requires calibrating the \I position when the star has been perfectly centred on the science fibre. This calibration could be completed in the laboratory during the AIT. Even so, the temporal stability of the opto-mechanical elements between the retro-injection and the science branches are uncertain, so there is a risk that the calibration needs to be redone regularly. Once the bundle is  installed at the telescope, it will not be possible to disconnect it easily to perform any calibration directly at the output of the science fibre. The recalibration of the exact position of the \S fibre with respect to \I will therefore require using data directly acquired with the spectrograph, which may be less straightforward than in the laboratory where a simple photometer can be used. This is nonetheless a possibility that will be carefully considered during commissioning.

The third strategy will hopefully combine the advantages of the first and the second strategies. It includes by design the compensation of the differential distortion using the retro injection and the \C fibre, and we will have a direct measurement of the flux injected into the \C fibre. On the other hand, the switch from the \C to the \S fibres for an off-axis object seems to be the limiting point in the first strategy. In the light of the performance study on MITHiC and considering the operational point of view, the first strategy is considered for HiRISE. However, it will be necessary to wait for the AIT to verify the strategy feasibility in the case of HiRISE.

The AIT phase of HiRISE in Europe is just starting. The alignment of the system will be based on a simplified SPHERE simulator that will reproduce the main characteristics of the SPHERE IFS beam using off-the-shelf components (LED source, lenses, and a pupil stop). This simulator will be used to align the bench and verify the image and wavefront quality. However, the simulator will be highly chromatic, which means that the system will be aligned with a monochromatic source at 1.6\,\mic and simply checked at other wavelengths. This implies that the centring strategies will not be fully testable during the AIT phase. Contrary to the MITHiC setup, HiRISE uses a dichroic filter to transmit all the science photons in $H$ band and uses the shorter wavelengths for the centring and calibrations. Some conclusions regarding the best centring strategy for HiRISE will come from the AIT phase, but the final decision will have to wait the first on-sky tests.

Due to the strong dependence of the performance to the global transmission of the system \citep{otten2021direct}, the centring procedures are crucial for HiRISE. In this particular case we work with existing systems, which implies strong constraints on the specifications in order to maximise the transmitted flux. These constraints will undoubtedly be relaxed for future instruments optimised end-to-end for HDC. Developing IFS instruments offering very high spectral resolution is challenging and costly, so it is likely that HDC will remain largely based on fibre-fed spectrographs in the foreseeable future. Our study therefore prepares the ground for instruments like RISTRETTO on the VLT \citep{Lovis2017} and, in the longer term, PCS on the ELT \citep{Kasper2021} or post-JWST space instruments.

\begin{acknowledgements}
    This project has received funding from the European Research Council (ERC) under the European Union’s Horizon 2020 research and innovation programme, grant agreements No. 757561 (HiRISE) and from Région Provence-Alpes-Côte d'Azur under grant agreement 2014-0276 (ASOREX project).
\end{acknowledgements}

\bibliographystyle{aa}
\bibliography{reference}


\end{document}